\documentclass[reqno]{amsart}

\makeatletter
\@addtoreset{equation}{section}

\makeatother

\newcommand{\Rm}{\mathbb{R}}
\newcommand{\Cm}{\mathbb{C}}

\newcommand{\Zm}{\mathbb{Z}}
\newcommand{\Sm}{\mathbb{S}}
\newcommand{\al}[1]{\begin{align}#1\end{align}}
\newcommand{\eq}[1]{\begin{align*}#1\end{align*}}
\newcommand{\va}{\varphi}
\newcommand{\pp}{\partial}
\newcommand{\vv}[1]{\boldsymbol{\mathrm{#1}}}
\newcommand{\hvv}[1]{\boldsymbol{\hat{\mathrm{#1}}}}
\newcommand{\uv}{\boldsymbol{{\hat{\mathrm{s}}}}}
\newcommand{\uvk}{\boldsymbol{\hat{\mathrm{k}}}}
\newcommand{\ket}[1]{\left|#1\right\rangle}
\newcommand{\bra}[1]{\left\langle#1\right|}
\newcommand{\braket}[2]{\left\langle#1\middle|#2\right\rangle}
\newcommand{\pint}{\:\mathcal{P}\!\!\int}
\newcommand{\rrf}[1]{\mathop{\mathcal{R}_{{#1}}}}
\newcommand{\irrf}[1]{\mathop{\mathcal{R}_{{#1}}^{-1}}}

\usepackage{amsmath}
\usepackage[foot]{amsaddr}
\usepackage{amsthm,amssymb,mathrsfs}
\usepackage{graphicx}
\usepackage{color}

\newtheorem{thm}{Theorem}[section]

\theoremstyle{remark}\newtheorem{rmk}[thm]{Remark}

\title[]{Numerical algorithms of the radiative transport equation using 
rotated reference frames for optical tomography with structured illumination}

\author[]{Manabu Machida}
\address{Institute for Medical Photonics Research, 
Hamamatsu University School of Medicine,
Hamamatsu 431-3192, Japan}
\email{machida@hama-med.ac.jp}

\date{\today}

\begin{document}

\begin{abstract}
We consider optical tomography with structured illumination in spatial-frequency domain using the three-dimensional radiative transport equation. Without the diffusion approximation, the radiative transport equation is solved by the technique of rotated reference frames. In addition to the method of rotated reference frames (spherical-harmonic expansion), the three dimensional $F_N$ method is applied to this optical tomography.
\end{abstract}

\maketitle

\section{Introduction}

Optical tomography is an imaging modality with near-infrared light \cite{Arridge99,Arridge-Hebden97,AS09,Gibson-etal05,Hebden-etal97}. Compared to the inverse problem of X-ray computed tomography, the inverse problem of optical tomography is more ill-posed since light is strongly scattered. Quite often arrays of optical fibers are used to detect outgoing light on the boundary. One way to improve the resolution of reconstructed images is to increase measured data. Noncontact optical tomography can readily acquire a large number of source-detector pairs \cite{AS09}. In a typical noncontact optical tomography, a source-detector pair consists of a point source by raster scanning a collimated laser beam and a pixel of a CCD camera \cite{Zhengmin05}. Since the energy density of light in random media such as biological tissue obeys the diffusion equation in the macroscopic regime, in which the propagation distance of light is much larger than the transport mean free path, optical tomography is usually formulated as an inverse problem of the diffusion equation. However, the diffusion approximation breaks in optically thin layers, near boundaries, and in strongly absorbing media. In this mesoscopic regime, in which the propagation distance of light is comparable to the transport mean free path, we need to use the radiative transport equation \cite{Case-Zweifel,Chandrasekhar60}, which has angular variables that do not exist in the diffusion equation.

In this paper, we consider noncontact optical tomography based on the radiative transport equation without making the diffusion approximation. We illuminate the boundary (the $x$-$y$ plane at $z=0$) of the half space in which the target inhomogeneity is embedded and measure the reflected light on the boundary. The Fourier transform is performed to the data from boundary measurements for source-detector pairs. By the use of spatially modulated beams of structured illumination \cite{Cuccia-etal05}, we can omit the Fourier transform for source positions. Thus the number of measurements can be reduced compared with raster scanned point sources. In this setup of optical tomography with structured illumination, improvement of spatial resolution was observed \cite{Bassi-etal09}. Absorbers of different structures can be reconstructed by using bi-dimensional source patterns \cite{DAndrea-etal10}. The use of angular-dependence of structured light reflectance \cite{Konecky-etal11} and dense sampling \cite{Zhou-etal13} were proposed. Albeit structured illumination is promising, optical tomography with structured illumination has been mostly limited to the inverse problem of the diffusion equation. Below we will develop transport-based optical tomography for structured illumination. In our optical tomography the absorption coefficient is recovered from boundary data directly measured in the spatial-frequency domain. A reasonably low-cost computation is achieved with algorithms using rotated reference frames.

In the half-space or slab geometry, it is known that plane-wave decomposition is useful \cite{AS09,Kaper69,Williams82}. Different numerical algorithms to compute the Green's function for the radiative transport equation as a sum of plane waves have been developed \cite{Barichello-Siewert00,Dunn-Siewert85,Kim04,Kim-Keller03,Kim-Moscoso04,Machida15a,Markel04,Panasyuk06,Siewert85,Siewert-Dunn83,Siewert-Dunn89,Williams09a,Williams09b}. In the case of isotropic scattering, in addition to \cite{Ganapol-etal94,Ganapol-Kornreich95,Ganapol-Kornreich09}, 
the three-dimensional radiative transport equation was considered by the pseudo-problem approach \cite{Williams82,Williams07} and the $F_N$ method \cite{Grandjean-Siewert79,Siewert78,Siewert-Benoist79}. By using discrete ordinates and plane-wave decomposition, Kim gave the Green's function as a sum of eigenmodes which are labeled by eigenvalues appearing in the corresponding one-dimensional problem \cite{Kim04}. Kim and his collaborators have applied the method to optical tomography \cite{Gonzalez-Rodriguez07,Gonzalez-Rodriguez09,Kim-Moscoso06}. Markel showed that eigenmodes in the Green's function with plane-wave decomposition are efficiently computed with the help of spherical-harmonic expansion in rotated reference frames \cite{Markel04}. Markel's method of rotated reference frames (MRRF) and has been intensively developed for the three-dimensional radiative transport equation in the half-space and slab geometry \cite{LK12a,LK12b,LK13a,LK13b,LK14,Machida10}. Optical tomography using the method of rotated reference frames was proposed \cite{Schotland-Markel07} and is verified by simulation and experiment \cite{Machida16a}. This method of rotated reference frames, however, sometimes suffers from numerical instability. The recently proposed three-dimensional $F_N$ method also uses plane-wave decomposition together with rotated reference frames \cite{Machida15a}.

Let $\vv{r}=(\vv{\rho},z)$ be a vector in $\Rm^3$, where $\vv{\rho}\in\Rm^2$ is a vector in the $x$-$y$ plane. We consider a medium occupying the half-space ($z>0$) in which light propagation is characterized by the absorption parameter $\mu_a$ and scattering parameter $\mu_s$. We assume that nonnegative $\mu_a$ depends on $\vv{r}$ and $\mu_s=\bar{\mu}_s$ is a positive constant. We write $\mu_a(\vv{r})$ as
\eq{
\mu_a(\vv{r})=\bar{\mu}_a+\delta\mu_a(\vv{r}),
}
where $\bar{\mu}_a$ is a constant. We introduce $\eta(\vv{r})$ as
\eq{
\eta(\vv{r})=\frac{\delta\mu_a(\vv{r})}{\bar{\mu}_t}
=(1-\varpi)\frac{\delta\mu_a(\vv{r})}{\bar{\mu}_a},
}
where $\varpi=\bar{\mu}_s/\bar{\mu}_t$ ($0<\varpi<1$) is the albedo for single scattering and
\eq{
\bar{\mu}_t=\bar{\mu}_a+\bar{\mu}_s.
}
We define
\eq{
\Rm^3_+=\left\{\vv{r}\in\Rm^3;\,z>0\right\},\quad
\Sm^2_{\pm}=\left\{\uv\in\Sm^2;\,\pm\mu>0\right\},
}
where $\mu$ is the cosine of the polar angle of $\uv$, and
\eq{
\Gamma_{\pm}=\left\{(\vv{r},\uv)\in\Rm^3\times\Sm^2_{\mp};\,z=0\right\}.
}
Let $I(\vv{r},\uv)$ ($\vv{r}\in\Rm^3_+$, $\uv\in\Sm^2$) be the specific intensity of light at position $\vv{r}$ traveling in direction $\uv$. We take the unit of length to be
\al{
\ell_t=\frac{1}{\bar{\mu}_t}.
\label{ellt}
}
The specific intensity $I(\vv{r},\uv)$ obeys the radiative transport equation,
\al{
\left\{\begin{aligned}
\uv\cdot\nabla I(\vv{r},\uv)+(1+\eta)I(\vv{r},\uv)
=\varpi\int_{\Sm^2}p(\uv,\uv')I(\vv{r},\uv')\,d\uv',
\\
(\vv{r},\uv)\in\Rm^3_+\times\Sm^2,
\\
I(\vv{r},\uv)=f(\vv{\rho},\uv),
\quad(\vv{r},\uv)\in\Gamma_-,
\end{aligned}\right.
\label{rte1}
}
where $\eta(\vv{r})$ is absorption inhomogeneity and $f(\vv{\rho},\uv)$ is the incident beam.  The scattering phase function $p(\uv,\uv')$ is normalized as
\eq{
\int_{\Sm^2}p(\uv',\uv)\,d\uv'=1,\qquad\uv\in\Sm^2.
}
Let $N$ be an integer. Assuming rotational symmetry we model $p(\uv,\uv')$ as
\eq{
p(\uv,\uv')
&=\frac{1}{4\pi}\sum_{l=0}^L\beta_lP_l(\uv\cdot\uv')
\\
&=\sum_{l=0}^L\sum_{m=-l}^l\frac{\beta_l}{2l+1}Y_{lm}(\uv)Y_{lm}^*(\uv'),
}
where $L\ge0$, $\beta_0=1$, $0<\beta_l<2l+1$ for $l\ge1$, and $P_l$ and $Y_{lm}$ are Legendre polynomials and spherical harmonics, respectively. Using associated Legendre polynomials $P_l^m$, $Y_{lm}$ are given by 
\eq{
Y_{lm}(\uv)=\sqrt{\frac{2l+1}{4\pi}\frac{(l-m)!}{(l+m)!}}P_l^m(\mu)e^{im\va}.
}
Here, $\va$ is the azimuthal angle of $\uv$. The symbol $*$ is used for complex conjugate. In the case of $\beta_l=(2l+1)\mathrm{g}^l$ and $L=\infty$, $p(\uv,\uv')$ is called the Henyey-Greenstein model \cite{HG41}. The constant $\mathrm{g}\in(-1,1)$ is called the scattering asymmetry parameter.

By using the radiative transport equation instead of the diffusion equation, we will generalize the noncontact diffuse optical tomography with structured illumination proposed by Lukic, Markel, and Schotland \cite{Lukic09}, which was also experimentally justified \cite{Konecky-etal09}. The remainder of this paper is organized as follows. We derive the linearized inverse problem in \S\ref{born}. \S\ref{pre} is devoted to the singular eigenfunctions and Green's function. In \S\ref{mrrf} and \S\ref{fn}, MRRF and the three-dimensional $F_N$ method are described. In \S\ref{mrrf}, we formulate MRRF by expanding the singular eigenfunctions with spherical harmonics. In \S\ref{ot} we begin by giving the spatially modulated beam and consider how we can reconstruct $\eta(\vv{r})$ after obtaining two specific intensities $I_s^{(i)}(\vv{r},\uv)$ ($i=1,2$). Numerical implementation is done in \S\ref{siml}. Finally in \S\ref{conclusions}, we make concluding remarks. Appendix \ref{slab} is devoted to the $F_N$ method for the slab geometry. The calculation of the forward data is presented in Appendix \ref{fwd}.

\section{Born approximation}
\label{born}

Let $I^{(0)}(\vv{r},\uv)$ be the specific intensity for $\eta(\vv{r})=0$, which obeys
\al{
\left\{\begin{aligned}
\left(\uv\cdot\nabla+1\right)I^{(0)}(\vv{r},\uv)
&=\varpi\int_{\Sm^2}p(\uv,\uv')I^{(0)}(\vv{r},\uv')\,d\uv',
\quad(\vv{r},\uv)\in\Rm^3_+\times\Sm^2,
\\
I^{(0)}(\vv{r},\uv)
&=
f(\vv{\rho},\uv),
\quad(\vv{r},\uv)\in\Gamma_-.
\end{aligned}\right.
\label{rte2}
}
We consider the Born series \cite{AS09},
\eq{
I(\vv{r},\uv)=I^{(0)}(\vv{r},\uv)-\int_{\Sm^2}\int_{\Rm^3_+}
G(\vv{r},\uv;\vv{r}',\uv')\eta(\vv{r}')I(\vv{r}',\uv')\,d\vv{r}'d\uv',
}
where the Green's function $G(\vv{r},\uv;\vv{r}_0,\uv_0)$ satisfies
\eq{
\left\{\begin{aligned}
\uv\cdot\nabla G(\vv{r},\uv;\vv{r}_0,\uv_0)
+G(\vv{r},\uv;\vv{r}_0,\uv_0)
&=
\varpi\int_{\Sm^2}p(\uv,\uv')G(\vv{r},\uv';\vv{r}_0,\uv_0)\,d\uv'
\\
&+
\delta(\vv{r}-\vv{r}_0)\delta(\uv-\uv_0),
\quad (\vv{r},\uv)\in\Rm^3_+\times\Sm^2,
\\
G(\vv{r},\uv;\vv{r}_0,\uv_0)
&=0,
\quad (\vv{r},\uv)\in\Gamma_-.
\end{aligned}\right.
}
If $\|\eta\|_{L^1(\Rm^3_+)}$ is sufficiently small, we can write
\al{
I(\vv{r},\uv)=
I^{(0)}(\vv{r},\uv)-
\int_{\Sm^2}\int_{\Rm^3_+}
G(\vv{r},\uv;\vv{r}',\uv')\eta(\vv{r}')I^{(0)}(\vv{r}',\uv')\,d\vv{r}'d\uv'.
\label{born:firstBorn}
}

Let us subtract the ballistic term by expressing $I^{(0)}(\vv{r},\uv)$ as
\eq{
I^{(0)}(\vv{r},\uv)=I_b(\vv{r},\uv)+I_s(\vv{r},\uv).
}
Here, $I_b(\vv{r},\uv)$ satisfies
\eq{
\left\{\begin{aligned}
\uv\cdot\nabla I_b(\vv{r},\uv)+I_b(\vv{r},\uv)=0,
&\quad (\vv{r},\uv)\in\Rm^3_+\times\Sm^2,
\\
I_b(\vv{r},\uv)=f(\vv{r},\uv),
&\quad (\vv{r},\uv)\in\Gamma_-,
\end{aligned}\right.
}
and $I_s(\vv{r},\uv)$ satisfies
\eq{
\left\{\begin{aligned}
\uv\cdot\nabla I_s(\vv{r},\uv)+I_s(\vv{r},\uv)
&=\varpi\int_{\Sm^2}p(\uv,\uv')I_s(\vv{r},\uv')\,d\uv'
+S[f](\vv{r},\uv),\quad (\vv{r},\uv)\in\Rm^3_+\times\Sm^2,
\\
I_s(\vv{r},\uv)&=0,
\quad (\vv{r},\uv)\in\Gamma_-.
\end{aligned}\right.
}
where
\eq{
S[f](\vv{r},\uv)=\varpi\int_{\Sm^2}p(\uv,\uv')I_b(\vv{r},\uv')\,d\uv'.
}
We have
\al{
I_b(\vv{r},\uv)
&=
\int_{\Sm^2_+}\int_{\Rm^3_+}\delta(\vv{\rho}-\vv{\rho}')
\frac{e^{-|z-z'|}}{\mu'}\delta(\uv-\uv')
\mu'f(\vv{\rho}',\uv')\delta(z')\,d\vv{r}'d\uv'
\nonumber \\
&=
e^{-z}f(\vv{\rho},\uv),
\label{ballisticIb}
}
for $\uv\in\Sm^2_+$, and $I_b=0$ for $\uv\in\Sm^2_-$. Hence
\eq{
S[f](\vv{r},\uv)=
\varpi\int_{\Sm^2}p(\uv,\uv')I_b(\vv{r},\uv')\,d\uv'=
\varpi e^{-z}\int_{\Sm^2_+}p(\uv,\uv')f(\vv{\rho},\uv')\,d\uv'.
}
Thus the rest is to compute the scattering term given by
\eq{
I_s(\vv{r},\uv)=\int_{\Sm^2}\int_{\Rm^3_+}G(\vv{r},\uv;\vv{r}',\uv')
S[f](\vv{r}',\uv')\,d\vv{r}'d\uv'.
}

\section{Preliminaries}
\label{pre}

Both of MRRF (the method of rotated reference frames) and the three-dimensional $F_N$ method are constructed from the three-dimensional Case's method \cite{Machida14}. Polynomials $g_l^m$ and $p_l^m$ are introduced in \S\ref{pre:poly}. Singular eigenfunctions are explained in \S\ref{case1d} and \S\ref{pre:case3d}. In \S\ref{pre:green}, the Green's function is given in terms of singular eigenfunctions.

\subsection{Polynomials}
\label{pre:poly}

Let us begin by introducing $h_l$ ($l=0,1,\dots$) as
\eq{
h_l=\left\{\begin{aligned}
2l+1-\varpi\beta_l,&\quad 0\le l\le L,
\\
2l+1,&\quad l>L.
\end{aligned}\right.
}

The normalized Chandrasekhar polynomials $g_l^m(\nu)$ ($m\ge0$, $l\ge m$, $\nu\in\Rm$) are given by the three-term recurrence relation \cite{Garcia-Siewert89,Garcia-Siewert90}
\eq{
\nu h_lg_l^m(\nu)
=\sqrt{(l+1)^2-m^2}g_{l+1}^m(\nu)+\sqrt{l^2-m^2}g_{l-1}^m(\nu),
}
with the initial term
\eq{
g_m^m(\nu)=\frac{(2m-1)!!}{\sqrt{(2m)!}}=\frac{\sqrt{(2m)!}}{2^{m}m!}.
}

Moreover we introduce the polynomials $p_l^m(\mu)$ ($m\ge0$, $l\ge m$) as
\eq{
p_l^m(\mu)
&=
(-1)^m\sqrt{\frac{(l-m)!}{(l+m)!}}P_l^m(\mu)(1-\mu^2)^{-m/2}
\\
&=
\sqrt{\frac{(l-m)!}{(l+m)!}}\frac{d^m}{d\mu^m}P_l(\mu).
}
The polynomials satisfy the three-term recurrence relation
\eq{
\sqrt{l^2-m^2}p_{l-1}^m(\mu)-(2l+1)\mu p_l^m(\mu)
+\sqrt{(l+1)^2-m^2}p_{l+1}^m(\mu)=0.
}

\subsection{Singular eigenfunctions for one dimension}
\label{case1d}

In one-dimensional transport theory, singular eigenfunctions $\phi^m(\nu,\mu)$ are given by \cite{Case60,McCormick-Kuscer66,Mika61}
\eq{
\phi^m(\nu,\mu)=\frac{\varpi\nu}{2}\mathcal{P}\frac{g^m(\nu,\mu)}{\nu-\mu}
+\lambda^m(\nu)\left(1-\mu^2\right)^{-|m|}\delta(\nu-\mu),
}
where $\mathcal{P}$ denotes the Cauchy principal value and 
\eq{
g^m(\nu,\mu)=\sum_{l=|m|}^L\beta_lp_l^m(\mu)g_l^m(\nu).
}
Here $|m|\le L$ and $\nu\in\Rm$ are eigenvalues; $\nu$ has discrete values $\pm\nu_j^m$ ($\nu_j^m>1$, $j=0,1,\dots,M^m-1$) and the continuous spectrum between $-1$ and $1$.  The number $M^m$ of discrete eigenvalues depends on $\varpi$ and $\beta_l$.  The function $\lambda^m(\nu)$ is given by
\eq{
\lambda^m(\nu)=1-\frac{\varpi\nu}{2}\pint_{-1}^1\frac{g^m(\nu,\mu)}{\nu-\mu}
(1-\mu^2)^{|m|}\,d\mu.
}
Singular eigenfunctions are normalized as
\eq{
\int_{-1}^1\phi^m(\nu,\mu)\left(1-\mu^2\right)^{|m|}\,d\mu=1.
}
We note that
\eq{
g_l^m(\nu)=(-1)^m\sqrt{\frac{(l-m)!}{(l+m)!}}
\int_{-1}^1\phi^m(\nu,\mu)(1-\mu^2)^{|m|/2}P_l^m(\mu)\,d\mu.
}
Discrete eigenvalues are roots of $\Lambda^m$, i.e., $\Lambda^m(\nu_j^m)=0$, where
\eq{
\Lambda^m(w)=1-\frac{\varpi w}{2}\int_{-1}^1\frac{g^m(w,\mu)}{w-\mu}
(1-\mu^2)^{|m|}\,d\mu.
}
We have the following orthogonality relations 
\cite{Case60,McCormick-Kuscer66,Mika61}
\eq{
\int_{-1}^1\mu\phi^m(\nu,\mu)\phi^m(\nu',\mu)(1-\mu^2)^{|m|}\,d\mu
=\mathcal{N}^m(\nu)\delta_{\nu\nu'},
}
where the Kronecker delta $\delta_{\nu\nu'}$ is replaced by the Dirac delta $\delta(\nu-\nu')$ if $\nu,\nu'$ are in the continuous spectrum. The normalization factor $\mathcal{N}^m(\nu)$ is given by
\eq{
\mathcal{N}^m(\nu)=\left\{\begin{aligned}
&\frac{1}{2}(\nu_j^m)^2g(\nu_j^m,\nu_j^m)\left.\frac{d\Lambda^m(w)}{dw}
\right|_{w=\nu_j^m},
\quad\nu=\nu_j^m,
\\
&\nu\Lambda^{m+}(\nu)\Lambda^{m-}(\nu)(1-\nu^2)^{-|m|},
\quad\nu\in(-1,1),
\end{aligned}\right.
}
where $\Lambda^{m\pm}(\nu)=\lim_{\epsilon\to0^+}\Lambda^m(\nu\pm i\epsilon)$.

Finally we introduce
\eq{
\Phi_{\nu}^m(\uv)=\phi^m(\nu,\mu)\left(1-\mu^2\right)^{|m|/2}e^{im\va}.
}

\subsection{Singular eigenfunctions for three dimensions}
\label{pre:case3d}

Let $\psi(\uv)\in\Cm$ be a function of $\uv\in\Sm^2$. By the operator $\rrf{\uvk}$ defined in \cite{Machida15a}, angles in $\rrf{\uvk}\psi(\uv)$ are measured in the rotated reference frame whose $z$-axis lies in the direction of a unit vector $\uvk\in\Cm^3$ ($\uvk\cdot\uvk=1$).

If $\psi(\uv)\in\Cm$ has the form
\eq{
\psi(\uv)=\sum_{l=0}^{\infty}\sum_{m=-l}^l\psi_{lm}Y_{lm}(\uv),
\qquad\psi_{lm}\in\Cm,
}
then we have \cite{Dede64,Kobayashi77,Markel04}
\eq{
\rrf{\uvk}\psi(\uv)
&=
\sum_{l=0}^{\infty}\sum_{m=-l}^l\psi_{lm}
\sum_{m'=-l}^le^{-im'\va_{\uvk}}d_{m'm}^l(\theta_{\uvk})Y_{lm'}(\uv),
\\
\irrf{\uvk}\psi(\uv)
&=
\sum_{l=0}^{\infty}\sum_{m=-l}^l\psi_{lm}
\sum_{m'=-l}^le^{im\va_{\uvk}}d_{mm'}^l(\theta_{\uvk})Y_{lm'}(\uv),
}
where $\theta_{\uvk}$ and $\va_{\uvk}$ are the polar and azimuthal angles of $\uvk$ in the laboratory frame, and $d_{m'm}^l$ are Wigner's $d$-matrices. In particular, we note that $\rrf{\uvk}\uv\cdot\uv'=\uv\cdot\uv'$, $\rrf{\uvk}\mu=\uv\cdot\uvk$, and
\eq{
\rrf{\uvk}Y_{lm}^*(\uv)
&=
(-1)^m\rrf{\uvk}Y_{l,-m}(\uv)
\\
&=
(-1)^m\sum_{m'=-l}^le^{-im'\va_{\uvk}}d_{m',-m}^l(\theta_{\uvk})Y_{lm'}(\uv)
\\
&=
\sum_{m'=-l}^l(-1)^{m'}e^{im'\va_{\uvk}}d_{m'm}^l(\theta_{\uvk})Y_{l,-m'}(\uv)
\\
&=
\sum_{m'=-l}^le^{im'\va_{\uvk}}d_{m'm}^l(\theta_{\uvk})Y_{lm'}^*(\uv).
}

\subsection{The Green's function}
\label{pre:green}

For the later purpose of the plane wave decomposition, we give complex unit vectors $\uvk(\nu,\vv{q})\in\Cm^3$ ($\nu\in\Rm$, $\vv{q}\in\Rm^2$) by
\eq{
\uvk(\nu,\vv{q})=\left(\begin{array}{c}-i\nu\vv{q}\\\hat{k}_z(\nu q)
\end{array}\right),
}
where $q=|\vv{q}|$ and
\eq{
\hat{k}_z(\nu q)=\sqrt{1+(\nu q)^2}.
}

It turns out that $d_{mm'}^l(\theta_{\uvk(\nu,\vv{q})})$ are functions of $\nu q$. We write
\eq{
d_{mm'}^l[i\tau(\nu q)]=d_{mm'}^l(\theta_{\uvk(\nu,\vv{q})}).
}
These $d$-matrices are computed using recurrence relations \cite{Machida10,Machida14,Machida15a}. To calculate $d_{mm'}^l[i\tau(\nu q)]$, we take square roots such that $0\le\mathop{\rm arg}(\sqrt{z})<\pi$ for all $z\in\Cm$ \cite{Panasyuk06,Machida10}.  We have
\eq{
\va_{\uvk(\nu,\vv{q})}=\left\{\begin{aligned}
\va_{\vv{q}}+\pi,&\quad\mbox{for}\,\nu>0,
\\
\va_{\vv{q}},&\quad\mbox{for}\,\nu<0,
\end{aligned}\right.
}
where $\va_{\vv{q}}$ is the polar angle of $\vv{q}$.

Let us consider the following homogeneous equation.
\eq{
\left(\uv\cdot\nabla+1\right)I(\vv{r},\uv)=
\varpi\int_{\Sm^2}p(\uv,\uv')I(\vv{r},\uv')\,d\uv'.
}
Let us consider the Fourier transform of the specific intensity:
\eq{
\tilde{I}(\vv{q},z,\uv)=
\int_{\Rm^2}e^{-i\vv{q}\cdot\vv{\rho}}I(\vv{r},\uv)\,d\vv{\rho}.
}
Then the Fourier transform $\tilde{I}(\vv{q},z,\uv)$ satisfies
\al{
\left(\mu\pp_z+i\vv{\omega}\cdot\vv{q}+1\right)\tilde{I}(\vv{q},z,\uv)
&=
\left(1-\frac{\uv\cdot\uvk}{\nu}\right)\tilde{I}(\vv{q},z,\uv)
\nonumber \\
&=
\varpi\int_{\Sm^2}p(\uv,\uv')\tilde{I}(\vv{q},z,\uv')\,d\uv',
\label{pre:green:homoeq}
}
where $\vv{\omega}\in\Rm^2$ was defined such that $\uv=(\vv{\omega},\mu)$. The solutions are given by
\eq{
\tilde{I}(\vv{q},z,\uv)=
e^{-\hat{k}_z(\nu q)z/\nu}\rrf{\uvk(\nu,\vv{q})}\Phi_{\nu}^m(\uv).
}

Three-dimensional singular eigenfunctions are obtained as
\eq{
\rrf{\uvk}\phi^m(\nu,\mu)=
\frac{\varpi\nu}{2}\mathcal{P}\frac{g^m(\nu,\uv\cdot\uvk)}{\nu-\uv\cdot\uvk}+
\lambda^m(\nu)\left(1-\nu^2\right)^{-|m|}\delta(\nu-\uv\cdot\uvk),
}
where $\uvk=\uvk(\nu,\vv{q})$.

The following orthogonality relation holds.
\eq{
\int_{\Sm^2}\mu\left(\rrf{\uvk(\nu,\vv{q})}\Phi_{\nu}^m(\uv)\right)
\left(\rrf{\uvk(\nu',\vv{q})}\Phi_{\nu'}^{m'*}(\uv)\right)\,d\uv
=
2\pi\hat{k}_z(\nu q)\mathcal{N}(\nu)\delta_{\nu\nu'}\delta_{mm'}.
}

Since the general solution is given by the sum of a particular solution and a linear combination of eigenmodes, we can write the Green's function as \cite{Case-Zweifel}
\eq{
G(\vv{r},\uv;\vv{r}_0,\uv_0)
&=
G_{\rm free}(\vv{r},\uv;\vv{r}_0,\uv_0)+
\frac{1}{(2\pi)^2}\int_{\Rm^2}e^{i\vv{q}\cdot(\vv{\rho}-\vv{\rho}_0)}
\\
&\times
\sum_{m=-L}^L\Biggl[\sum_{j=0}^{M^m-1}\hat{A}^m(\nu_j^m)
\rrf{\uvk(\nu_j^m,\vv{q})}\Phi_j^m(\uv)e^{-\hat{k}_z(\nu_j^mq)z/\nu_j^m}
\\
&+
\int_0^1\hat{A}^m(\nu)\rrf{\uvk(\nu,\vv{q})}\Phi_{\nu}^m(\uv)
e^{-\hat{k}_z(\nu q)z/\nu}\,d\nu\Biggr]\,d\vv{q},
}
where $G_{\rm free}(\vv{r},\uv;\vv{r}_0,\uv_0)$ is the free-space Green's function and $\hat{A}^m(\nu_j^m,\vv{q})$, $\hat{A}^m(\nu,\vv{q})$ are some coefficients which are determined from the boundary conditions. We note that the free-space Green's function or the fundamental solution is obtained as \cite{Machida14}
\eq{
G_{\rm free}(\vv{r},\uv;\vv{r}_0,\uv_0)=
\frac{1}{(2\pi)^2}\int_{\Rm^2}e^{i\vv{q}\cdot(\vv{\rho}-\vv{\rho}_0)}
\tilde{G}_{\rm free}(z,\uv;z_0,\uv_0;\vv{q})\,d\vv{q},
}
where
\eq{
&
\tilde{G}_{\rm free}(z,\uv;z_0,\uv_0;\vv{q})
\\
&=
\sum_{m=-L}^L\Biggl[\sum_{j=0}^{M^m-1}\frac{1}{2\pi\hat{k}_z(\nu_j^mq)
\mathcal{N}(\nu_j^m)}
\rrf{\uvk(\nu_j^m,\vv{q})}\Phi_{j\pm}^m(\uv)\Phi_{j\pm}^{m*}(\uv_0)
e^{-\hat{k}_z(\nu_j^mq)|z-z_0|/\nu_j^m}
\\
&+
\int_0^1\frac{1}{2\pi\hat{k}_z(\nu q)\mathcal{N}(\nu)}
\rrf{\uvk(\pm\nu,\vv{q})}
\Phi_{\pm\nu}^m(\uv)\Phi_{\pm\nu}^{m*}(\uv_0)
e^{-\hat{k}_z(\nu q)|z-z_0|/\nu}\,d\nu\Biggr].
}
Upper signs are chosen for $z>z_0$ and lower signs are chosen for $z<z_0$. The Fourier transform of $G(\vv{r},\uv;\vv{r}_0,\uv_0)$ is defined similarly.

Let us consider the Fourier transform $\tilde{I}_s$ of $I_s$, which is given by
\eq{
I_s(\vv{r},\uv)=\frac{1}{(2\pi)^2}\int_{\Rm^2}
e^{i\vv{q}\cdot\vv{\rho}}\tilde{I}_s(\vv{q},z,\uv)\,d\vv{q}.
}
We can express $\tilde{I}_s(\vv{q},z,\uv)$ in terms of the Green's function and singular eigenfunctions as
\al{
&
\tilde{I}_s(\vv{q},z,\uv)
=\int_0^{\infty}\int_{\Sm^2}\tilde{G}(z,\uv;z',\uv';\vv{q})
S[\tilde{f}](\vv{q},z',\uv')\,d\uv'dz'
\nonumber \\
&=
\int_0^{\infty}\int_{\Sm^2}\tilde{G}_{\rm free}(z,\uv;z',\uv';\vv{q})
S[\tilde{f}](\vv{q},z',\uv')\,d\uv'dz'
\nonumber \\
&+
\sum_{m=-L}^L\Biggl[\sum_{j=0}^{M^m-1}
A^m(\nu_j^m)\rrf{\uvk(\nu_j^m,\vv{q})}\Phi_j^m(\uv)
e^{-\hat{k}_z(\nu_j^mq)z/\nu_j^m}
\nonumber \\
&+
\int_0^1A^m(\nu)\rrf{\uvk(\nu,\vv{q})}\Phi_{\nu}^m(\uv)
e^{-\hat{k}_z(\nu q)z/\nu}\,d\nu\Biggr],
\label{expressionIs}
}
with some coefficients $A^m(\nu_j^m,\vv{q})$, $A^m(\nu,\vv{q})$. We note that $\tilde{S}[f]=S[\tilde{f}]$, where $\tilde{S}(\vv{q},z,\uv)=\int_{\Rm^2}e^{-i\vv{q}\cdot\vv{\rho}}S(\vv{r},\uv)\,d\vv{\rho}$ and $\tilde{f}(\vv{q},\uv)=\int_{\Rm^2}e^{-i\vv{q}\cdot\vv{\rho}}f(\vv{\rho},\uv)\,d\vv{\rho}$.

\section{Method of rotated reference frames}
\label{mrrf}

The method of rotated reference frames was first proposed by Markel \cite{Markel04}. Here, we formulate the method making the relation to Case's method clear.

In the method of rotated reference frames, we expand the singular eigenfunction as
\eq{
\Phi_{\nu}^m(\uv)=\sum_{l=|m|}^NC_l^m(\nu)Y_{lm}(\uv).
}
By multiplying (\ref{pre:green:homoeq}) by $\rrf{\uvk}Y_{l_1m_1}^*(\uv)$ and integrating both sides, we obtain
\eq{
&
\int_{\Sm^2}\left[\rrf{\uvk}Y_{l_1m_1}^*(\uv)\right]
\left(1-\frac{\uv\cdot\uvk}{\nu}\right)\sum_{l=0}^N
C_l^m(\nu)\rrf{\uvk}Y_{lm}(\uv)\,d\uv
\\
&=
\varpi\int_{\Sm^2}\int_{\Sm^2}p(\uv,\uv')
\left[\rrf{\uvk}Y_{l_1m_1}^*(\uv)\right]\sum_{l=0}^NC_l^m(\nu)
\rrf{\uvk}Y_{lm}(\uv')\,d\uv'd\uv.
}
Hence,
\eq{
&
C_{l_1}^m(\nu)-\frac{1}{\nu}\sum_{l_2=0}^NC_{l_2}^m(\nu)
\int_{\Sm^2}\mu Y_{l_1m}^*(\uv)Y_{l_2m}(\uv)\,d\uv
\\
&=
\varpi\sum_{l_2=0}^NC_{l_2}^m(\nu)\sum_{l'=0}^L\sum_{m'=-l'}^{l'}
\frac{\beta_{l'}}{2l'+1}
\int_{\Sm^2}\int_{\Sm^2}Y_{l'm'}(\uv)Y_{l'm'}^*(\uv')Y_{l_1m}^*(\uv)
Y_{l_2m}(\uv')\,d\uv d\uv'.
}
Therefore,
\eq{
\frac{1}{\nu}\sum_{l_2=0}^NR_{l_1l_2}^m
C_{l_2}^m(\nu)=\frac{h_{l_1}}{2l_1+1}C_{l_1}^m(\nu),
}
where
\eq{
R_{l_1l_2}^m=\int_{\Sm^2}\mu Y_{l_1m}^*(\uv)Y_{l_2m}(\uv)\,d\uv.
}
Let us define vector $\ket{\psi_{\nu}(m)}$ and matrix $B(m)$ whose components and entries are given by
\eq{
\braket{l}{\psi_{\nu}(m)}&=
\frac{1}{\sqrt{Z_{\nu}(m)}}\sqrt{\frac{h_l}{2l+1}}C_l^m(\nu),
\\
B_{ll'}(m)&=\sqrt{\frac{(2l+1)(2l'+1)}{h_lh_{l'}}}R_{ll'}^m.
}
The normalization factor $Z_{\nu}(m)$ is calculated below. Thus we have \cite{Machida10,Panasyuk06}
\eq{
B(m)\ket{\psi_{\nu}(m)}=\nu\ket{\psi_{\nu}(m)}.
}
The tridiagonal matrix $B(m)$ is given by
\eq{
B_{ll'}(m)=\sqrt{\frac{l^2-m^2}{h_{l-1}h_l}}\delta_{l',l-1}+
\sqrt{\frac{(l+1)^2-m^2}{h_lh_{l+1}}}\delta_{l',l+1}.
}
Since the eigenvalues $\nu$ depend on $m$, we can write $\nu=\nu_j^m$. In numerical calculation we introduce $l_B$ ($\ge L$, $-L\le m\le L$) and write the matrix $B(m)$ as
\al{
B(m)=\left(\begin{array}{ccccc}
0&b_{|m|+1}&0&&\\
b_{|m|+1}&0&b_{|m|+2}&&\\
0&b_{|m|+2}&0&\ddots&\\
&&\ddots&\ddots&b_{l_B}\\
&&&b_{l_B}&0
\end{array}\right),
\label{mrrf:Bmat}
}
where $b_l(m)=\sqrt{(l^2-m^2)/(h_lh_{l-1})}$.  The matrix $B(m)$ has $(l_B-|m|+1)/2$ or $(l_B-|m|)/2$ positive eigenvalues for $l_B-|m|+1$ even or odd, respectively. 

We determine the normalization constant $Z_{\nu}(m)$ from the condition $\braket{\psi_{\nu}(m)}{\psi_{\nu}(m)}=1$. We have
\eq{
&
\int_{\Sm^2}\mu\left|\Phi_{\nu}^m(\uv)\right|^2\,d\uv
\\
&=
\int_{\Sm^2}\mu\sum_{l_1=|m|}^N\sum_{l_2=|m|}^NC_{l_1}^m(\nu)C_{l_2}^{m*}(\nu)
Y_{l_1m}(\uv)Y_{l_2m}^*(\uv)\,d\uv
\\
&=
\sum_{l_1=|m|}^N\left[C_{l_1}^m(\nu)C_{l_1-1}^{m*}(\nu)
\sqrt{\frac{l_1^2-m^2}{4l_1^2-1}}
+C_{l_1}^m(\nu)C_{l_1+1}^{m*}(\nu)\sqrt{\frac{(l_1+1)^2-m^2}{4(l_1+1)^2-1}}
\right]
\\
&=
Z_{\nu}(m)
\sum_{l_1=|m|}^N\left[b_{l_1}(m)\braket{\psi_{\nu}(m)}{l_1-1}+
b_{l_1+1}(m)\braket{\psi_{\nu}(m)}{l_1+1}\right]
\braket{l_1}{\psi_{\nu}(m)}
\\
&=
Z_{\nu}(m)\sum_{l_1=|m|}^N\sum_{l_2=|m|}^N\braket{\psi_{\nu}(m)}{l_2}
B_{l_1l_2}(m)\braket{l_1}{\psi_{\nu}(m)}
\\
&=
Z_{\nu}(m)\bra{\psi_{\nu}(m)}B(m)\ket{\psi_{\nu}(m)}^*
\\
&=
Z_{\nu}(m)\nu\braket{\psi_{\nu}(m)}{\psi_{\nu}(m)}.
}
On the other hand,
\eq{
\int_{\Sm^2}\mu\left|\Phi_{\nu}^m(\uv)\right|^2\,d\uv=
2\pi\mathcal{N}^m(\nu).
}
Therefore we obtain \cite{Machida14}
\eq{
Z_{\nu}(m)=\frac{2\pi\mathcal{N}^m(\nu)}{\nu}.
}

In this way, the three-dimensional singular eigenfunction is given in the context of MRRF as
\eq{
\rrf{\uvk(\nu,\vv{q})}\Phi_{\nu}^m(\uv)=
\sqrt{\frac{2\pi\mathcal{N}^m(\nu)}{\nu}}\sum_{l=|m|}^N
\sqrt{\frac{2l+1}{h_l}}\braket{l}{\psi_{\nu}(m)}
\rrf{\uvk(\nu,\vv{q})}Y_{lm}(\uv).
}

Since the three-dimensional singular eigenfunctions are expressed as a superposition of $\rrf{\uvk(\nu,\vv{q})}Y_{lm}(\uv)$ in MRRF, we can obtain $\tilde{I}_s(\vv{q},z,\uv)$ in (\ref{expressionIs}) without calculating the fundamental solution $\tilde{G}_{\rm free}(z,\uv;z_0,\uv_0;\vv{q})$.

Let us introduce $I_{M\nu}^{(\pm)}(\vv{r},\uv;\vv{q})$ as \cite{Panasyuk06,Machida10}
\eq{
I_{M\nu}^{(+)}(\vv{r},\uv;\vv{q})
&=
e^{i\vv{q}\cdot\vv{\rho}-\hat{k}_z(\nu q)z/\nu}\sum_{l=0}^N
\sqrt{\frac{2l+1}{h_l}}\sum_{m=-l}^lY_{lm}(\uv)
\\
&\times
(-1)^me^{-im\va_{\vv{q}}}\braket{l}{\psi_{\nu}(M)}d_{mM}^l[i\tau(\nu q)],
\\
I_{M\nu}^{(-)}(\vv{r},\uv;\vv{q})
&=
e^{i\vv{q}\cdot\vv{\rho}+\hat{k}_z(\nu q)z/\nu}\sum_{l=0}^N
\sqrt{\frac{2l+1}{h_l}}\sum_{m=-l}^lY_{lm}(-\uv)
\\
&\times
e^{-im\va_{\vv{q}}}\braket{l}{\psi_{\nu}(M)}d_{m,-M}^l[i\tau(\nu q)].
}
Then we have
\eq{
I^{(0)}(\vv{r},\uv)=\frac{1}{(2\pi)^2}\sum_{M=-L}^L\sum_{\nu>0}
\int_{\Rm^2}F_{M\nu}^{(+)}(\vv{q})I_{M\nu}^{(+)}(\vv{r},\uv;\vv{q})\,d\vv{q},
}
where $F_{M\nu}^{(+)}(\vv{q})$ are determined from the boundary condition and $\sum_{\nu>0}$ stands for the sum over all positive eigenvalues of $B(M)$. From the boundary condition we obtain on $\Gamma_-$,
\eq{
&
\sum_{M=-L}^L\sum_{\nu>0}\sum_{l'=0}^N\sqrt{\frac{2l'+1}{h_{l'}}}
\sum_{m'=-l'}^{l'}Y_{l'm'}(\uv)(-1)^{m'}\braket{l'}{\psi_{\nu}(M)}
\\
&\times
d_{m'M}^{l'}[i\tau(\nu q)]e^{-im'\va_{\vv{q}}}
F_{M\nu}^{(+)}(\vv{q})=\tilde{f}(\vv{q},\uv).
}
For later purpose we consider
\al{
f^{(1)}(\vv{r},\uv)=\delta(\vv{\rho}),\quad
f^{(2)}(\vv{r},\uv)=\delta(\vv{\rho})\delta(\uv-\hvv{z}),
\label{f1f2}
}
and write $\tilde{f}(\vv{q},\uv)=\tilde{f}^{(i)}(\vv{q},\uv)$ ($i=1,2$), where
\al{
\begin{aligned}
\tilde{f}^{(1)}(\vv{q},\uv)&=1,
\\
\tilde{f}^{(2)}(\vv{q},\uv)&=\delta(\uv-\hvv{z})
=\sum_{l=0}^N\sqrt{\frac{2l+1}{4\pi}}Y_{l0}(\uv),
\end{aligned}
\label{ft1ft2}
}
where we used $Y_{lm}(\hvv{z})=\sqrt{(2l+1)/(4\pi)}\delta_{m0}$. Correspondingly we write $F_{M\nu}^{(+)}(\vv{q})=F_{M\nu}^{(+)i}(\vv{q})$ ($i=1,2$). By multiplying $Y_{lm}^*(\uv)$ on both sides and integrating in $\Sm^2_+$, we have
\eq{
&
\sum_{M=-L}^L\sum_{\nu>0}\sum_{l'=|m|}^N\sqrt{\frac{2l'+1}{h_{l'}}}
\mathcal{B}_{ll'}^m\braket{l'}{\psi_{\nu}(M)}d_{mM}^{l'}[i\tau(\nu q)]
\\
&\times
\begin{vmatrix}F_{M\nu}^{(+)1}(\vv{q}) \\ F_{M\nu}^{(+)2}(\vv{q})\end{vmatrix}
=
\delta_{m0}
\begin{vmatrix}
\sqrt{4\pi}\mathcal{B}_{l0}^0 \\
\sum_{l'=0}^N\sqrt{\frac{2l'+1}{4\pi}}\mathcal{B}_{ll'}^0
\end{vmatrix}.
}
Here we defined
\eq{
\mathcal{B}_{ll'}^m
&=
\int_{\Sm^2_+}Y_{lm}^*(\uv)Y_{l'm}(\uv)\,d\uv
\\
&=
\frac{1}{2}\sqrt{\frac{(2l+1)(2l'+1)(l-m)!(l'-m)!}{(l+m)!(l'+m)!}}
\int_0^1P_l^m(\mu)P_{l'}^m(\mu)\,d\mu,
}
and used
\eq{
\int_0^1P_l(\mu)\,d\mu
&=
\frac{2}{\sqrt{2l+1}}\mathcal{B}_{l0}^0
\\
&=
\left\{\begin{aligned}
1,&\quad l=0,\\
0,&\quad\mbox{even}\,l\,(\neq0),\\
(-1)^{(l-1)/2}\frac{l!!}{l(l+1)(l-1)!!},&\quad\mbox{odd}\,l.
\end{aligned}\right.
}
The fact that $(-1)^MF_{-M,\nu}^{(+)}$ satisfies the same equation as $F_{M\nu}^{(+)}$ implies
\eq{
(-1)^MF_{-M,\nu}^{(+)}=F_{M\nu}^{(+)},
}
where we note that \cite{Markel04} $\braket{l}{\psi_{-\nu}(M)}=(-1)^l\braket{l}{\psi_{\nu}(M)}$. We obtain
\eq{
&
\sum_{M=0}^L\sum_{\nu>0}\sum_{l'=|m|}^N\sqrt{\frac{2l'+1}{h_{l'}}}
\mathcal{B}_{ll'}^m\braket{l'}{\psi_{\nu}(M)}
\\
&\times
\left[
d_{mM}^{l'}[i\tau(\nu q)]+(1-\delta_{M0})(-1)^Md_{m,-M}^{l'}[i\tau(\nu q)]
\right]
\\
&\times
\begin{vmatrix}F_{M\nu}^{(+)1}(\vv{q}) \\ F_{M\nu}^{(+)2}(\vv{q})\end{vmatrix}
=
\delta_{m0}
\begin{vmatrix}
\sqrt{4\pi}\mathcal{B}_{l0}^0 \\
\sum_{l'=0}^N\sqrt{\frac{2l'+1}{4\pi}}\mathcal{B}_{ll'}^0
\end{vmatrix}.
}
Thus $F_{M\nu}^{(+)i}(\vv{q})$ ($0\le M\le N$) are obtained from the linear system,
\eq{
\mathcal{M}(q)\vv{F}^{(+)i}=\vv{v}^{(+)i},\quad i=1,2,
}
where the matrix $\mathcal{M}(q)$ is defined as
\eq{
\{\mathcal{M}(q)\}_{lm,M\nu}
&=
\sum_{l'=|m|}^N\sqrt{\frac{2l'+1}{h_{l'}}}\mathcal{B}_{ll'}^m
\braket{l'}{\psi_{\nu}(M)}
\\
&\times
\left[d_{mM}^{l'}[i\tau(\nu q)]+
(1-\delta_{M0})(-1)^Md_{m,-M}^{l'}[i\tau(\nu q)]\right],
}
and vectors $\vv{F}^{(+)i},\vv{v}^{(+)i}$ are given by
\eq{
\{\vv{F}^{(+)i}\}_{M\nu}
&=
F_{M\nu}^{(+)i},\quad i=1,2,
\\
\{\vv{v}^{(+)1}\}_{lm}
&=
\delta_{m0}\sqrt{4\pi}\mathcal{B}_{l0}^0,
\\
\{\vv{v}^{(+)2}\}_{lm}
&=
\delta_{m0}\sum_{l'=0}^N\sqrt{\frac{2l'+1}{4\pi}}\mathcal{B}_{ll'}^0.
}
The matrix $\mathcal{M}(q)$ becomes square when $L=M$, $l=|m|+2\alpha-1$, $\alpha=1,\dots,\lfloor(N-|m|+1)/2\rfloor$ (see Remark \ref{rmk1} below). That is,
\eq{
I^{(0)}(\vv{r},\uv)
&=
\frac{1}{(2\pi)^2}\int_{\Rm^2}e^{i\vv{q}\cdot\vv{\rho}}
\sum_{M=-L}^L\sum_{\nu>0}e^{-\hat{k}_z(\nu q)z/\nu}F_{M\nu}^{(+)i}(\vv{q})
\\
&\times
\sum_{l=0}^N\sum_{m=-l}^l\sqrt{\frac{2l+1}{h_l}}
\braket{l}{\psi_{\nu}(M)}d_{mM}^l[i\tau(\nu q)]
\\
&\times
(-1)^me^{-im\va_{\vv{q}}}Y_{lm}(\uv)\,d\vv{q},\quad i=1,2.
}

Now, we note that the ballistic term (\ref{ballisticIb}) is calculated as
\eq{
I_b(\vv{r},\uv)
&=
e^{-z}
\begin{vmatrix}
f^{(1)}(\vv{\rho},\uv)\\f^{(2)}(\vv{\rho},\uv)
\end{vmatrix}
\\
&=
\frac{1}{(2\pi)^2}\int_{\Rm^2}e^{i\vv{q}\cdot\vv{\rho}}e^{-z}
\begin{vmatrix}
\sqrt{4\pi}Y_{00}(\uv) \\ \sum_{l=0}^N\sqrt{\frac{2l+1}{4\pi}}Y_{l0}(\uv)
\end{vmatrix}
\,d\vv{q}.
}
Therefore,
\eq{
I_s(\vv{r},\uv)
&=
I^{(0)}(\vv{r},\uv)-I_b(\vv{r},\uv)
\\
&=
\frac{1}{(2\pi)^2}\int_{\Rm^2}e^{i\vv{q}\cdot\vv{\rho}}
\sum_{m=-N}^N\sum_{l=|m|}^N\begin{vmatrix}
\gamma_{lm}^{(1)}(\vv{q},z) \\ \gamma_{lm}^{(2)}(\vv{q},z)
\end{vmatrix}Y_{lm}(\uv)\,d\vv{q}.
}
Here we defined
\eq{
\gamma_{lm}^{(1)}(\vv{q},z)
&=
\sqrt{\frac{2l+1}{h_l}}(-1)^me^{-im\va_{\vv{q}}}\hat{\gamma}_{lm}^{(1)}(q,z)
-\sqrt{4\pi}\delta_{l0}\delta_{m0}e^{-z},
\\
\gamma_{lm}^{(2)}(\vv{q},z)
&=
\sqrt{\frac{2l+1}{h_l}}(-1)^me^{-im\va_{\vv{q}}}\hat{\gamma}_{lm}^{(2)}(q,z)
-\delta_{m0}\sqrt{\frac{2l+1}{4\pi}}e^{-z},
}
where
\eq{
\hat{\gamma}_{lm}^{(i)}(q,z)
&=
\sum_{M=-L}^L\sum_{\nu>0}F_{M\nu}^{(+)i}(\vv{q})
e^{-\hat{k}_z(\nu q)z/\nu}\braket{l}{\psi_{\nu}(M)}
d_{mM}^l[i\tau(\nu q)],\quad i=1,2.
}

\section{Three-dimensional $F_N$ method}
\label{fn}

By combining the $F_N$ method and the technique of rotated reference frames, we can establish a numerical scheme for the three-dimensional radiative transport equation with anisotropic scattering.

Let us introduce the notation
\eq{
\sideset{}{'}\sum_{lm}
=\sum_{m=-N}^N\sum_{\alpha=0\atop l=|m|+2\alpha}^{\lfloor(N-|m|)/2\rfloor}.
}
For $\uv\in\Sm^2_+$, we expand $\tilde{I}_s$ as
\eq{
\tilde{I}_s(\vv{q},0,-\uv)&=
\sideset{}{'}\sum_{lm}c_{lm}(\vv{q})Y_{lm}(\uv),
\\
\tilde{I}_s(\vv{q},z,-\uv)&=
\sideset{}{'}\sum_{lm}b_{lm}(\vv{q},z)Y_{lm}(\uv),
\\
\tilde{I}_s(\vv{q},z,\uv)&=
\sideset{}{'}\sum_{lm}a_{lm}(\vv{q},z)Y_{lm}(\uv).
}

We define $\xi_j^m$ ($-L\le m\le L$) as
\al{
\xi_j^m=
\left\{\begin{aligned}
\nu_j^m,&\quad j=0,1,\dots,M^m-1,
\\
\cos\left(\frac{\pi}{2}\frac{j-M^m+1}{N_{\rm row}^m-M^m+1}\right),
&\quad j=M^m,\dots,N_{\rm row}^m-1,
\end{aligned}\right.
\nonumber \\
\label{defxi}
}
where
\al{
N_{\rm row}^m
=\left\lfloor\frac{N-|m|}{2}\right\rfloor+1.
}
We drop the superscript $m$ if there is no confusion.

\begin{rmk}
\label{rmk1}
How to discretize $\nu$ in the continuous spectrum depends on $N,L$. We selected $\xi_j^m$ in (\ref{defxi}) because we set $N=L$ in the numerical calculation below.
\end{rmk}

Furthermore we introduce
\eq{
\mathcal{I}_{lm}^{m'}(\xi_j,\vv{q})
&=
\int_{\Sm^2}\left(\rrf{\uvk(\xi_j,\vv{q})}\Phi_{\xi_j}^{m'*}(\uv)\right)
Y_{lm}(\uv)\,d\uv.
\\
&=
\sqrt{(2l+1)\pi}(-1)^{m'}e^{im\va_{\uvk}}d_{mm'}^l[i\tau(\xi_jq)]
g_l^{m'}(\xi_j),
}
where $\mathcal{I}_{lm}^{m'}(\xi_j,\vv{q})=0$ for $|m'|>l$. We have
\al{
\int_{\Sm^2}\mu\left(\rrf{\uvk}\Phi_{-\xi_j}^{m'*}(\uv)\right)
\tilde{I}_s(\vv{q},0,\uv)\,d\uv
&=
L_1^{m'}[\tilde{f}](-\xi_j,\vv{q},0),
\label{fn:proj1}
\\
\int_{\Sm^2}\mu\left(\rrf{\uvk}\Phi_{\xi_j}^{m'*}(\uv)\right)
\tilde{I}_s(\vv{q},0,\uv)\,d\uv
&=
2\pi\hat{k}_z(\xi_jq)\mathcal{N}^{m'}(\xi_j)A^{m'}(\xi_j),
\label{fn:proj3}
\\
\int_{\Sm^2}\mu\left(\rrf{\uvk}\Phi_{-\xi_j}^{m'*}(\uv)\right)
\tilde{I}_s(\vv{q},z,\uv)\,d\uv
&=
L_1^{m'}[\tilde{f}](-\xi_j,\vv{q},z),
\label{fn:proj5}
\\
\int_{\Sm^2}\mu\left(\rrf{\uvk}\Phi_{\xi_j}^{m'*}(\uv)\right)
\tilde{I}_s(\vv{q},z,\uv)\,d\uv
&=
2\pi\hat{k}_z(\xi_jq)\mathcal{N}^{m'}(\xi_j)
A^{m'}(\xi_j)e^{-\hat{k}_z(\xi_jq)z/\xi_j}
\nonumber \\
&+
L_2^{m'}[\tilde{f}](\xi_j,\vv{q},z),
\label{fn:proj6}
}
where
\eq{
&
L_1^{m'}[\tilde{f}](-\xi_j,\vv{q},z)
\\
&=\int_z^{\infty}\int_{\Sm^2}
\left(\rrf{\uvk(-\xi_j,\vv{q})}\Phi_{-\xi_j}^{m'*}(\uv')\right)
e^{-\hat{k}_z(\xi_jq)(z'-z)/\xi_j}
\tilde{S}[\tilde{f}](\vv{q},z',\uv')\,d\uv'dz'
\\
&=
e^{-z}\frac{\varpi\xi_j}{\xi_j+\hat{k}_z(\xi_jq)}\sum_{l=0}^L\sum_{m=-l}^l
\frac{\beta_l}{2l+1}
\mathcal{I}_{lm}^{m'}(-\xi_j,\vv{q})
\int_{\Sm^2_+}\tilde{f}(\vv{q},\uv)Y_{lm}^*(\uv)\,d\uv,
}
\eq{
&
L_2^{m'}[\tilde{f}](\xi_j,\vv{q},z)
\\
&=
\int_0^z\int_{\Sm^2}
\left(\rrf{\uvk(\xi_j,\vv{q})}\Phi_{\xi_j}^{m'*}(\uv')\right)
e^{-\hat{k}_z(\xi_jq)(z-z')/\xi_j}
\tilde{S}[\tilde{f}](\vv{q},z',\uv')\,d\uv'dz'
\\
&=
\left(e^{-\hat{k}_z(\xi_jq)z/\xi_j}-e^{-z}\right)
\frac{\varpi\xi_j}{\xi_j-\hat{k}_z(\xi_jq)}\sum_{l=0}^L\sum_{m=-l}^l
\frac{\beta_l}{2l+1}
\mathcal{I}_{lm}^{m'}(\xi_j,\vv{q})
\int_{\Sm^2_+}\tilde{f}(\vv{q},\uv)Y_{lm}^*(\uv)\,d\uv.
}
In particular for $\tilde{f}^{(i)}(\vv{q},\uv)$ ($i=1,2$) in (\ref{ft1ft2}), we have
\eq{
&
L_1^{m'}[\tilde{f}](-\xi_j,\vv{q},z)
=e^{-z}\frac{\varpi\xi_j}{2(\xi_j+\hat{k}_z(\xi_jq))}
\sum_{l=|m'|}^L(-1)^l\beta_l
\\
&\times
d_{0m'}^l[i\tau(\xi_jq)]g_l^{m'}(\xi_j)
\left\{\begin{aligned}
\frac{4\pi}{\sqrt{2l+1}}\mathcal{B}_{l0}^0,&\quad i=1,
\\
1,&\quad i=2,
\end{aligned}\right.
}
\eq{
&
L_2^{m'}[\tilde{f}](\xi_j,\vv{q},z)
=\left(e^{-\hat{k}_z(\xi_jq)z/\xi_j}-e^{-z}\right)
\frac{\varpi\xi_j}{2(\xi_j-\hat{k}_z(\xi_jq))}
\\
&\times
\sum_{l=|m'|}^L\beta_ld_{0m'}^l[i\tau(\xi_jq)]g_l^{m'}(\xi_j)
\left\{\begin{aligned}
\frac{4\pi}{\sqrt{2l+1}}\mathcal{B}_{l0}^0,&\quad i=1,
\\
1,&\quad i=2.
\end{aligned}\right.
}
By noticing $g_l^{-m}(\nu)=(-1)^mg_l^m(\nu)$, we see that $L_1^{m'}[\tilde{f}]$ and $L_2^{m'}[\tilde{f}]$ are independent of $\va_{\vv{q}}$ and independent of the sign of $m'$.

From (\ref{fn:proj1}) we obtain
\eq{
-(-1)^{m'}\int_{\Sm^2_+}\mu\left(\rrf{\uvk}\Phi_{\xi_j}^{m'*}(\uv)\right)
\tilde{I}_s(\vv{q},0,-\uv)\,d\uv
=
L_1^{m'}[\tilde{f}](-\xi_j,\vv{q},0),
}
where we used $\Phi_{-\xi_j}^{m'}(\uv)=(-1)^{m'}\Phi_{\xi_j}^{m'}(-\uv)$. We introduce
\eq{
&
\mathcal{J}_{lm}^{(\pm)jm'}
=
e^{-im\va_{\vv{q}}}\int_{\Sm^2_{\pm}}\mu\left(\rrf{\uvk(-\xi_j,\vv{q})}
\Phi_{-\xi_j}^{m'*}(-\uv)\right)Y_{lm}(\uv)\,d\uv
\\
&=
(-1)^{l+1}e^{-im\va_{\vv{q}}}\int_{\Sm^2_{\mp}}\mu
\left(\rrf{\uvk(-\xi_j,\vv{q})}
\Phi_{-\xi_j}^{m'*}(\uv)\right)Y_{lm}(\uv)\,d\uv.
}
Explicit expressions of $\mathcal{J}_{lm}^{(\pm)jm'}$ are found as follows \cite{Machida15a}.
\eq{
&
\mathcal{J}_{lm}^{(+)jm'}
=\hat{k}_z(\xi_jq)\sqrt{\frac{\pi}{2l+1}}(-1)^md_{mm'}^l[i\tau(\xi_jq)]
\\
&\times
\left(\sqrt{(l+1)^2-{m'}^2}g_{l+1}^{m'}(\xi_j)
+\sqrt{l^2-{m'}^2}g_{l-1}^{m'}(\xi_j)\right)
\\
&
-i\frac{|\xi_jq|}{2}\sqrt{\frac{\pi}{2l+1}}(-1)^m
\sum_{m''=-l}^ld_{mm''}^l[i\tau(\xi_jq)]
\\
&\times
\Biggl[\delta_{m'',m'-1}\Bigl(\sqrt{(l-m'')(l-m')}g_{l-1}^{m'}(\xi_j)-
\sqrt{(l+m'+1)(l+m')}g_{l+1}^{m'}(\xi_j)\Bigr)
\\
&+
\delta_{m'',m'+1}\Bigl(\sqrt{(l-m'+1)(l-m')}g_{l+1}^{m'}(\xi_j)-
\sqrt{(l+m'')(l+m')}g_{l-1}^{m'}(\xi_j)\Bigr)\Biggr]
\\
&-
\mathcal{J}_{lm}^{(-)jm'},
}
where we used $g_l^m(-\nu)=(-1)^{l+m}g_l^m(\nu)$, and
\eq{
&
\mathcal{J}_{lm}^{(-)jm'}
=
\frac{\varpi\xi_j}{2}(-1)^{l+1}\sqrt{\frac{2l+1}{4\pi}\frac{(l-m)!}{(l+m)!}}
[\mathop{\rm sgn}(m')]^{m'}
\\
&\times
\frac{\sqrt{(2|m'|)!}}{(2|m'|-1)!!}
\sum_{m''=-|m'|}^{|m'|}(-1)^{m''}\sqrt{\frac{(|m'|-m'')!}{(|m'|+m'')!}}
d_{m'',-m'}^{|m'|}[i\tau(\xi_jq)]
\\
&\times
\int_{\Sm^2_+}\frac{g^{m'}\left(-\xi_j,\hat{k}_z(\xi_jq)\mu+i\xi_jq
\sqrt{1-\mu^2}\cos\va\right)}{\xi_j+\hat{k}_z(\xi_jq)\mu+i\xi_jq
\sqrt{1-\mu^2}\cos\va}
\mu P_{|m'|}^{m''}(\mu)P_l^m(\mu)e^{i(m+m'')\va}\,d\uv.
}
When numerically evaluating the above integral over $\uv$, we use the Gauss-Legendre quadrature for $\mu$ and trapezoidal rule for $\va$. We note that
\al{
\mathcal{J}_{l,-m}^{(\pm)j,-m'}=(-1)^m\mathcal{J}_{lm}^{(\pm)jm'}.
\label{Jplusminus}
}

For each $\xi_j,\vv{q}$, we have
\eq{
\sideset{}{'}\sum_{lm}\mathcal{J}_{lm}^{(+)jm'}c_{lm}(\vv{q})e^{im\va_{\vv{q}}}
=-L_1^{m'}[\tilde{f}](-\xi_j,\vv{q},0).
}
Using (\ref{Jplusminus}) and the fact that $L_1^{-m'}=L_1^{m'}$, we have
\eq{
&
\sideset{}{'}\sum_{lm}\mathcal{J}_{lm}^{(+)jm'}c_{lm}(\vv{q})e^{im\va_{\vv{q}}}
\\
&=
\sideset{}{'}\sum_{lm}\mathcal{J}_{lm}^{(+)j,-m'}
c_{lm}(\vv{q})e^{im\va_{\vv{q}}}
\\
&=
\sideset{}{'}\sum_{lm}\mathcal{J}_{l,-m}^{(+)j,-m'}
c_{l,-m}(\vv{q})e^{-im\va_{\vv{q}}}
\\
&=
\sideset{}{'}\sum_{lm}\mathcal{J}_{lm}^{(+)jm'}
c_{l,-m}(\vv{q})(-1)^me^{-im\va_{\vv{q}}}.
}
The above relation implies
\eq{
c_{lm}(\vv{q})=\hat{c}_{lm}(q)e^{-im\va_{\vv{q}}},\quad
\hat{c}_{l,-m}(q)=(-1)^m\hat{c}_{lm}(q).
}
That is,
\eq{
&
\sum_{m=0}^N\sum_{l=|m|,|m|+2,\dots}\left[\mathcal{J}_{lm}^{(+)jm'}
+(1-\delta_{m0})(-1)^m\mathcal{J}_{l,-m}^{(+)jm'}\right]
\hat{c}_{lm}(q)=-L_1^{m'}[\tilde{f}](-\xi_j,\vv{q},0).
}

We obtain from (\ref{fn:proj3}) and (\ref{fn:proj6})
\al{
&
\int_{\Sm^2}\mu\left(\rrf{\uvk}\Phi_{\xi_j}^{m'*}(\uv)\right)
\tilde{I}_s(\vv{q},z,\uv)\,d\uv
\nonumber \\
&=
e^{-\hat{k}_z(\xi_jq)z/\xi_j}\int_{\Sm^2}\mu
\left(\rrf{\uvk}\Phi_{\xi_j}^{m'*}(\uv)\right)\tilde{I}_s(\vv{q},0,\uv)\,d\uv
+L_2^{m'}[\tilde{f}](\xi_j,\vv{q},z).
\label{fn:key}
}
Together with (\ref{fn:proj5}), we have
\eq{
&
\sideset{}{'}\sum_{lm}(-1)^{l+1}\mathcal{J}_{lm}^{(-)jm'}
a_{lm}(\vv{q},z)e^{im\va_{\vv{q}}}
\\
&=
\sideset{}{'}\sum_{lm}\mathcal{J}_{lm}^{(+)jm'}
b_{lm}(\vv{q},z)e^{im\va_{\vv{q}}}
+L_1^{m'}[\tilde{f}](-\xi_j,\vv{q},z),
\\
&
\sideset{}{'}\sum_{lm}(-1)^{m'}\mathcal{J}_{lm}^{(+)jm'}
a_{lm}(\vv{q},z)e^{im\va_{\vv{q}}}-
\sideset{}{'}\sum_{lm}(-1)^{l+m'+1}\mathcal{J}_{lm}^{(-)jm'}
b_{lm}(\vv{q},z)e^{im\va_{\vv{q}}}
\\
&=
e^{-\hat{k}_z(\xi_jq)z/\xi_j}(-1)^{m'}\left(\sideset{}{'}\sum_{lm}(-1)^l
\mathcal{J}_{lm}^{(-)jm'}c_{lm}(\vv{q})e^{im\va_{\vv{q}}}\right)+
L_2^{m'}[\tilde{f}](\xi_j,\vv{q},z),
}
where we used $Y_{lm}(-\uv)=(-1)^lY_{lm}(\uv)$. Let us define
\eq{
v_1^{jm'}
&=
-L_1^{m'}[\tilde{f}](-\xi_j,\vv{q},z),
\\
v_2^{jm'}
&=
e^{-\hat{k}_z(\xi_jq)z/\xi_j}\left(\sideset{}{'}\sum_{lm}(-1)^l
\mathcal{J}_{lm}^{(-)jm'}c_{lm}(q)\right)
\\
&+
(-1)^{m'}L_2^{m'}[\tilde{f}](\xi_j,\vv{q},z).
}
We can show that $v_1^{jm'}$ and $v_2^{jm'}$ are independent of $\va_{\vv{q}}$ 
and satisfy $v_1^{j,-m'}=v_1^{jm'}$ and $v_2^{j,-m'}=v_2^{jm'}$. Similarly to 
$c_{lm}(\vv{q})$ we have
\eq{
a_{lm}(\vv{q},z)&=\hat{a}_{lm}(q,z)e^{-im\va_{\vv{q}}},
\\
\hat{a}_{l,-m}(q,z)&=(-1)^m\hat{a}_{lm}(q,z),
\\
b_{lm}(\vv{q},z)&=\hat{b}_{lm}(q,z)e^{-im\va_{\vv{q}}},
\\
\hat{b}_{l,-m}(q,z)&=(-1)^m\hat{b}_{lm}(q,z).
}
We write
\eq{
&
\mathcal{M}_{jm',lm}^{(11)}=\mathcal{M}_{jm',lm}^{(22)}=
(-1)^l\mathcal{J}_{lm}^{(-)jm'}
+(1-\delta_{m0})(-1)^{l+m}\mathcal{J}_{l,-m}^{(-)jm'},
\\
&
\mathcal{M}_{jm',lm}^{(12)}=\mathcal{M}_{jm',lm}^{(21)}=
\mathcal{J}_{lm}^{(+)jm'}+(1-\delta_{m0})(-1)^m\mathcal{J}_{l,-m}^{(+)jm'}.
}
We obtain
\[
\left(\begin{array}{c|c}
\mathcal{M}_{jm',lm}^{(11)}&\mathcal{M}_{jm',lm}^{(12)}\\
\hline
\mathcal{M}_{jm',lm}^{(21)}&\mathcal{M}_{jm',lm}^{(22)}
\end{array}\right)
\left(\begin{array}{c}
\hat{a}_{lm}(q,z)\\
\hline
\hat{b}_{lm}(q,z)
\end{array}\right)
=\left(\begin{array}{c}
v_1^{jm'}\\
\hline
v_2^{jm'}
\end{array}\right).
\]
By setting $N=L$ and choosing $0\le m'\le L$, $0\le j\le N_{\rm row}^{m'}-1$, 
$0\le m\le N$, $l=m+2\alpha$, $0\le\alpha\le\left\lfloor\frac{N-m}{2}\right\rfloor$, the above matrix $\mathcal{M}$ becomes a square matrix. Otherwise the above linear system can be solved by singular value decomposition.

In Appendix \ref{slab}, the three-dimensional $F_N$ method for the slab geometry is explained.

\section{Optical tomography with structured illumination}
\label{ot}

We here explain the set up of our optical tomography. The data function introduced in \S\ref{Dfunc} is obtained for spatially modulated incident beams in \S\ref{ot:st}, and $\eta(\vv{r})$ is reconstructed according to the inversion formula (\ref{ot:reconst:reconstructedeta}) in \S\ref{ot:reconst}.

\subsection{Structured illumination}
\label{ot:st}

Let us consider structured illumination in the half space, i.e., the incoming beam is given by
\eq{
f(\vv{\rho},\uv)
&=
F_{B_0}(\vv{\rho},\uv)
\\
&=
I_0\left[1+A_0\cos(\vv{q}_0\cdot\vv{\rho}+B_0)\right]\delta(\uv-\uv_0),
}
where $\mu_0\in(0,1]$ is the cosine of the polar angle of $\uv_0$, $I_0$ is the amplitude, $A_0$ is the modulation depth, and $B_0$ is the phase of the source. Since 
\eq{
\frac{2F_0(\vv{\rho},\uv)-
\left(1-i\sqrt{3}\right)F_{-2\pi/3}(\vv{\rho},\uv)
-\left(1+i\sqrt{3}\right)F_{2\pi/3}(\vv{\rho},\uv)
}{3A_0}=
f_{\vv{q}_0}(\vv{\rho},\uv),
}
where
\eq{
f_{\vv{q}_0}(\vv{\rho},\uv)=I_0e^{i\vv{q}_0\cdot\vv{\rho}}\delta(\uv-\uv_0),
\quad\mu_0\in(0,1],
}
we will use $f(\vv{\rho},\uv)=f_{\vv{q}_0}(\vv{\rho},\uv)$ for the boundary condition \cite{Lukic09}. Then we have
\al{
\tilde{f}(\vv{q},\uv)
&=\int_{\Rm^2}e^{-i\vv{q}\cdot\vv{\rho}}f(\vv{\rho},\uv)\,d\vv{\rho}
\nonumber \\
&=
(2\pi)^2I_0\delta(\vv{q}-\vv{q}_0)\delta(\uv-\uv_0).
\label{fourierf}
}
We measure the exitance or hemispheric flux $J_+$ defined as follows on the boundary.
\eq{
J_+(\vv{\rho})=\int_{\Sm^2_+}\mu I(\vv{\rho},0,-\uv)\,d\uv.
}

Moreover we consider the Fourier transform of $\eta$:
\eq{
\tilde{\eta}(\vv{q},z)
=\int_{\Rm^2}e^{-i\vv{q}\cdot\vv{\rho}}\eta(\vv{\rho},z)\,d\vv{\rho}.
}
Since the reconstructed $\tilde{\eta}$ is regularized, the reconstructed $\eta$ is given by
\al{
\eta(\vv{\rho},z)
&=
\frac{1}{(2\pi)^2}\int_{\Rm^2}e^{i\vv{q}\cdot\vv{\rho}}
\chi_{\Omega_B}(\vv{q})\tilde{\eta}(\vv{q},z)\,d\vv{q}
\nonumber \\
&=:
\eta_B(\vv{\rho},z),
\label{etaB}
}
where $\Omega_B$ is a subdomain in the first Brillouin zone (see \S\ref{Dfunc}) and $\chi_{\Omega_B}(\vv{q})$ is the characteristic function such that $\chi_{\Omega_B}(\vv{q})=1$ for $\vv{q}\in\Omega_B$ and $\chi_{\Omega_B}(\vv{q})=0$ otherwise.

\subsection{The data function}
\label{Dfunc}

We refer to $D(\vv{\rho})$ below as the data function.
\eq{
D(\vv{\rho})
&=
J_+^{(0)}(\vv{\rho})-J_+(\vv{\rho})
\\
&=
\int_{\Sm^2_+}\mu\left[I^{(0)}(\vv{\rho},0,-\uv)-I(\vv{\rho},0,-\uv)\right]
\,d\uv.
}
Within the first Born approximation, the data function $D(\vv{\rho})$ is given by
\al{
D(\vv{\rho})=
\int_{\Sm^2}\int_{\Rm^3_+}\left[\int_{\Sm^2_+}\mu
G(\vv{\rho},0,-\uv;\vv{r}',\uv')\,d\uv\right]
\eta_B(\vv{r}')I^{(0)}(\vv{r}',\uv')\,d\vv{r}'d\uv'.
\label{datafunc}
}
We note that $D(\vv{\rho})$ is obtained through the exitance measured on the boundary. We can reconstruct $\eta_B$ by solving the linear inverse problem in (\ref{datafunc}).

Suppose that there are $(2N_d+1)\times(2N_d+1)$ detectors on grid points $\vv{\rho}=(x_i,y_j)$ ($i,j=-N_d,\dots,N_d$) with spacing $h_d$. We consider the Fourier transform
\eq{
\tilde{D}(\vv{q})=\sum_{\vv{\rho}}e^{-i\vv{q}\cdot\vv{\rho}}D(\vv{\rho}).
}
Noting the Poisson summation formula
\eq{
\sum_{\vv{\rho}}e^{i\vv{q}\cdot\vv{\rho}}
=\left(\frac{2\pi}{h_d}\right)^2\sum_{\vv{p}}\delta(\vv{q}+\vv{p}),
} 
where $\vv{p}$ denotes reciprocal lattice points $(2\pi i/h_d,2\pi j/h_d)$ 
($i,j\in\Zm$). We note that $\Omega_B$ is in the the first Brillouin zone 
$[-\pi/h_d,\pi/h_d]\times[-\pi/h_d,\pi/h_d]$. If $\vv{q}-\vv{q}_0\in\Omega_B$, we obtain
\al{
\tilde{D}(\vv{q})
&=
\frac{\mu_0I_0}{h_d^2}\int_0^{\infty}\int_{\Sm^2}\left[
\int_{\Sm^2_+}\mu\tilde{G}^*(z',-\uv';0,\uv;\vv{q})\,d\uv\right]
\nonumber \\
&\times
\tilde{\eta}_B(\vv{q}-\vv{q}_0,z')
\tilde{G}(z',\uv';0,\uv_0;\vv{q}_0)\,d\uv'dz'.
\label{ot:DGG}
}
To compute the right-hand side of (\ref{ot:DGG}), we recall $f^{(1)},f^{(2)}$ in (\ref{f1f2}) and introduce
\eq{
I^{(i)}(\vv{r},\uv)
=\int_{\Sm^2_+}\int_{\Rm^3_+}G(\vv{r},\uv;\vv{r}',\uv')
\mu'f^{(i)}(\vv{\rho}',\uv')\delta(z')\,d\vv{r}'d\uv',
}
for $i=1,2$ and its Fourier transform
\eq{
\tilde{I}^{(i)}(\vv{q},z,\uv)=\int_{\Sm^2_+}
\tilde{G}(z,\uv;0,\uv';\vv{q})\mu'\tilde{f}^{(i)}(\vv{q},\uv')\,d\uv',
}
for $i=1,2$. We note that $\tilde{f}^{(i)}$ are given in (\ref{ft1ft2}) and $I^{(i)}(\vv{r},\uv)$ satisfies (\ref{rte2}) with the boundary source $f^{(i)}(\vv{\rho},\uv)$ ($i=1,2$). Thus we can write $\tilde{D}(\vv{q})$ as
\al{
\tilde{D}(\vv{q})=
\frac{I_0}{h_d^2}\int_0^{\infty}\int_{\Sm^2}\tilde{I}^{(1)*}(\vv{q},z',-\uv')
\tilde{\eta}_B(\vv{q}-\vv{q}_0,z')
\tilde{I}^{(2)}(\vv{q}_0,z',\uv')\,d\uv'dz'.
\label{DI1I2}
}

By redefining $\vv{q}$ as $\vv{q}-\vv{q}_0\rightarrow\vv{q}$ in (\ref{DI1I2}), we have
\al{
\mathcal{D}(\vv{q}_0,\vv{q})
=\int_0^{\infty}K(\vv{q}_0,z;\vv{q})\tilde{\eta}_B(\vv{q},z)\,dz,\qquad
\vv{q}\in\Omega_B,
\label{ot:reconst:DMeta}
}
where
\eq{
\mathcal{D}(\vv{q}_0,\vv{q})=\tilde{D}(\vv{q}+\vv{q}_0)\frac{h_d^2}{I_0},
}
and
\eq{
K(\vv{q}_0,z;\vv{q})
=\int_{\Sm^2}\tilde{I}^{(1)*}(\vv{q}+\vv{q}_0,z,-\uv)
\tilde{I}^{(2)}(\vv{q}_0,z,\uv)\,d\uv.
}
We obtain
\al{
K(\vv{q}_0,z;\vv{q})
&=
e^{-z}\int_{\Sm^2_+}\tilde{I}_s^{(1)*}(\vv{q}+\vv{q}_0,z,-\uv)
\tilde{f}^{(2)}(\vv{q}_0,\uv)\,d\uv
\nonumber \\
&+
e^{-z}\int_{\Sm^2_+}\tilde{I}_s^{(2)}(\vv{q}_0,z,-\uv)
\tilde{f}^{(1)*}(\vv{q}+\vv{q}_0,\uv)\,d\uv
\nonumber \\
&+
\int_{\Sm^2}\tilde{I}_s^{(1)*}(\vv{q}+\vv{q}_0,z,-\uv)
\tilde{I}_s^{(2)}(\vv{q}_0,z,\uv)\,d\uv.
\label{Kexpress}
}
Thus the quality of reconstruction of $\eta$ in this inverse problem is determined by the kernel $K$ in (\ref{ot:reconst:DMeta}). So far in most research including \cite{Lukic09}, $K$ has been computed within the diffusion approximation. In the present paper, we will calculate $K$ directly from the radiative transport equation.

Below, we will numerically compute $\tilde{I}_s^{(1)},\tilde{I}_s^{(2)}$ using MRRF and the three-dimensional $F_N$ method.

\subsection{Reconstruction by MRRF}
\label{ot:reconst}

We expand $\tilde{I}_s^{(1)}$ and $\tilde{I}_s^{(2)}$ with spherical harmonics:
\eq{
\tilde{I}_s^{(1)}(\vv{q},z',-\uv')
&=
\sum_{m=-N}^N\sum_{l=|m|}^N\gamma_{lm}^{(1)}(\vv{q},z')Y_{lm}(-\uv'),
\quad \uv'\in\Sm^2,
\\
\tilde{I}_s^{(2)}(\vv{q}_0,z',\uv')
&=
\sum_{m=-N}^N\sum_{l=|m|}^N\gamma_{lm}^{(2)}(\vv{q}_0,z')Y_{lm}(\uv'),
\quad \uv'\in\Sm^2.
}
Using MRRF, we obtain
\al{
K(\vv{q}_0,z;\vv{q})
&=
e^{-z}\sum_{l=0}^N\sum_{l'=0}^N(-1)^l\gamma_{l0}^{(1)*}(\vv{q}+\vv{q}_0,z)
\mathcal{B}_{ll'}^0+
e^{-z}\sqrt{4\pi}\sum_{l=0}^N(-1)^l\gamma_{l0}^{(2)}(\vv{q}_0,z)
\mathcal{B}_{l0}^0
\nonumber \\
&+
\sum_{m=-N}^N\sum_{l=|m|}^N(-1)^l\gamma_{lm}^{(1)*}(\vv{q}+\vv{q}_0,z)
\gamma_{lm}^{(2)}(\vv{q}_0,z),
\label{ot:reconst:Kab}
}
where we used $Y_{lm}(-\uv)=(-1)^lY_{lm}(\uv)$ and $\int_{\Sm^2_+}Y_{lm}(\uv)\,d\uv=\sqrt{4\pi}\delta_{m0}\mathcal{B}_{l0}^0$. Hence,
\eq{
&
K(\vv{q}_0,z;\vv{q})
\\
&=
e^{-z}\sum_{l=0}^N\sum_{l'=0}^N(-1)^l
\left[\sqrt{\frac{2l+1}{h_l}}\hat{\gamma}_{l0}^{(1)*}(|\vv{q}+\vv{q}_0|,z)
-e^{-z}\sqrt{4\pi}\delta_{l0}\right]\mathcal{B}_{ll'}^0
\\
&+
e^{-z}\sqrt{4\pi}\sum_{l=0}^N(-1)^l\left[\sqrt{\frac{2l+1}{h_l}}
\hat{\gamma}_{l0}^{(2)}(q_0,z)-e^{-z}\sqrt{\frac{2l+1}{4\pi}}\right]
\mathcal{B}_{l0}^0
\\
&+
\sum_{l=0}^N(-1)^l\left[\sqrt{\frac{2l+1}{h_l}}
\hat{\gamma}_{l0}^{(1)*}(|\vv{q}+\vv{q}_0|,z)-e^{-z}\sqrt{4\pi}\delta_{l0}
\right]
\\
&\times
\left[\sqrt{\frac{2l+1}{h_l}}\hat{\gamma}_{l0}^{(2)}(q_0,z)
-e^{-z}\sqrt{\frac{2l+1}{4\pi}}\right]
\\
&+
2\sum_{m=1}^N\sum_{l=m}^N(-1)^l\frac{2l+1}{h_l}
\hat{\gamma}_{lm}^{(1)*}(|\vv{q}+\vv{q}_0|,z)
\hat{\gamma}_{lm}^{(2)}(q_0,z)
\cos\left(m(\va_{\vv{q}+\vv{q}_0}-\va_{\vv{q}_0})\right).
}

Viewing (\ref{ot:reconst:DMeta}) as a linear matrix-vector equation we can express (\ref{ot:reconst:DMeta}) as
\eq{
\ket{\mathcal{D}(\vv{q})}=\hat{K}(\vv{q})\ket{\tilde{\eta}_B(\vv{q})},
}
where $\braket{\vv{q}_0}{\mathcal{D}(\vv{q})}=\mathcal{D}(\vv{q}_0,\vv{q})$, $\bra{\vv{q}_0}\hat{K}(\vv{q})\ket{z}=K(\vv{q}_0,z;\vv{q})$, and $\braket{z}{\tilde{\eta}_B(\vv{q})}=\tilde{\eta}_B(\vv{q},z)$. We can compute $\ket{\tilde{\eta}_B(\vv{q})}$ with singular value decomposition. We obtain
\eq{
\tilde{\eta}_B(\vv{q},z)=
\bra{z}\hat{K}^+(\vv{q})\ket{\mathcal{D}(\vv{q})},
}
where $\hat{K}^+$ is the pseudoinverse such that
\eq{
\hat{K}^+=
\hat{K}^{\dagger}\left(\hat{K}\hat{K}^{\dagger}\right)_{\rm reg}^{-1}.
}
Here $\dagger$ denotes the Hermitian conjugate and ${\rm reg}$ means that the pesudoinverse is regularized as is explained below. Let $\sigma_j^2(\vv{q})$ and $\ket{v_j(\vv{q})}$ be the eigenvalues and eigenvectors of the matrix $\hat{M}(\vv{q})$ whose $\vv{q}_0$-$\vv{q}'_0$ element is given by
\eq{
\bra{\vv{q}_0}\hat{M}(\vv{q})\ket{\vv{q}'_0}
=\int_0^{\infty}K(\vv{q}_0,z;\vv{q})K^*(\vv{q}'_0,z;\vv{q})
\,dz.
}
If we use the truncated SVD and take only singular values greater than a threshold value $\sigma_0$ as regularization, $\eta$ is reconstructed as
\al{
\eta(\vv{r})
&=
\frac{1}{(2\pi)^2}\int_{\Omega_B}e^{i\vv{q}\cdot\vv{\rho}}
\sum_{j\atop\sigma_j>\sigma_0}\sigma_j(\vv{q})^{-2}\sum_{\vv{q}_0}
\braket{v_j(\vv{q})}{\vv{q}_0}
\nonumber \\
&\times
\mathcal{D}(\vv{q}_0,\vv{q})\sum_{\vv{q}'_0}
K^*(\vv{q}'_0,z;\vv{q})\braket{\vv{q}'_0}{v_j(\vv{q})}\,d\vv{q}.
\label{ot:reconst:reconstructedeta}
}
The resulting reconstruction is shown in Fig.~\ref{fig1}.

\subsection{Reconstruction by the three-dimensional $F_N$ method}
\label{ot:reconstfn}

We expand $\tilde{I}_s^{(1)}$ and $\tilde{I}_s^{(2)}$ with spherical harmonics:
\eq{
\tilde{I}_s^{(1)}(\vv{q},z',-\uv')=
\left\{\begin{aligned}
\sum_{m=-N}^N\sum_{\alpha=0\atop l=|m|+2\alpha}^{\lfloor(N-|m|)/2\rfloor}
b_{lm}^{(1)}(\vv{q},z')Y_{lm}(\uv'),
\quad \uv'\in\Sm^2_+,
\\
\sum_{m=-N}^N\sum_{\alpha=0\atop l=|m|+2\alpha}^{\lfloor(N-|m|)/2\rfloor}
a_{lm}^{(1)}(\vv{q},z')Y_{lm}(-\uv'),
\quad \uv'\in\Sm^2_-,
\end{aligned}\right.
}
and
\eq{
\tilde{I}_s^{(2)}(\vv{q}_0,z',\uv')=
\left\{\begin{aligned}
\sum_{m=-N}^N\sum_{\alpha=0\atop l=|m|+2\alpha}^{\lfloor(N-|m|)/2\rfloor}
a_{lm}^{(2)}(\vv{q}_0,z')Y_{lm}(\uv'),
\quad \uv'\in\Sm^2_+,
\\
\sum_{m=-N}^N\sum_{\alpha=0\atop l=|m|+2\alpha}^{\lfloor(N-|m|)/2\rfloor}
b_{lm}^{(2)}(\vv{q}_0,z')Y_{lm}(-\uv'),
\quad \uv'\in\Sm^2_-.
\end{aligned}\right.
}
Note that $Y_{lm}(-\uv')=(-1)^lY_{lm}(\uv')$. Since $\uv_0\in\Sm^2_+$ and $-\uv\in\Sm^2_-$ in (\ref{ot:DGG}), the ballistic terms do not contribute to $\tilde{D}(\vv{q})$. The kernel in (\ref{Kexpress}) is obtained as
\al{
&
K(\vv{q}_0,z;\vv{q})
\nonumber \\
&=
e^{-z}\sum_{m=-N}^N\sum_{\alpha=0\atop l=|m|+2\alpha}^{\lfloor(N-|m|)/2\rfloor}
b_{lm}^{(1)*}(\vv{q}+\vv{q}_0,z)Y_{lm}^*(\hvv{z})+
e^{-z}\sqrt{4\pi}\sum_{\alpha=0\atop l=2\alpha}^{\lfloor N/2\rfloor}
b_{l0}^{(2)}(\vv{q}_0,z)\mathcal{B}_{l0}^0
\nonumber \\
&+
\sum_{m=-N}^N\sum_{\alpha=0\atop l=|m|+2\alpha}^{\lfloor(N-|m|)/2\rfloor}
\sum_{\alpha'=0\atop l'=|m|+2\alpha'}^{\lfloor(N-|m|)/2\rfloor}
\Bigl[a_{lm}^{(2)}(\vv{q}_0,z)b_{l'm}^{(1)*}(\vv{q}+\vv{q}_0,z)
\nonumber \\
&+
b_{lm}^{(2)}(\vv{q}_0,z)a_{l'm}^{(1)*}(\vv{q}+\vv{q}_0,z)\Bigr]
\mathcal{B}_{ll'}^m.
\label{ot:reconstfn:Kab}
}
Thus the kernel $K$ is obtained using the three-dimensional $F_N$ method. We can rewrite (\ref{ot:reconstfn:Kab}) as
\eq{
&
K(\vv{q}_0,z;\vv{q})=
e^{-z}\sum_{\alpha=0\atop l=2\alpha}^{\lfloor N/2\rfloor}
\sqrt{\frac{2l+1}{4\pi}}\hat{b}_{l0}^{(1)*}(|\vv{q}+\vv{q}_0|,z)+
e^{-z}\sqrt{4\pi}\sum_{\alpha=0\atop l=2\alpha}^{\lfloor N/2\rfloor}
\hat{b}_{l0}^{(2)}(q_0,z)\mathcal{B}_{l0}^0
\\
&+
\sum_{m=0}^N\sum_{\alpha=0\atop l=|m|+2\alpha}^{\lfloor(N-m)/2\rfloor}
\sum_{\alpha'=0\atop l'=|m|+2\alpha'}^{\lfloor(N-m)/2\rfloor}
\left[\hat{a}_{lm}^{(2)}(q_0,z)\hat{b}_{l'm}^{(1)*}(|\vv{q}+\vv{q}_0|,z)+
\hat{b}_{lm}^{(2)}(q_0,z)\hat{a}_{l'm}^{(1)*}(|\vv{q}+\vv{q}_0|,z)\right]
\\
&\times
\left[\delta_{m0}+2(1-\delta_{m0})\cos\left(m(\va_{\vv{q}+\vv{q}_0}
-\va_{\vv{q}_0})\right)\right]\mathcal{B}_{ll'}^m,
}
where we used $(-1)^mY_{l,-m}^*(\uv)=Y_{lm}(\uv)$. The absorption inhomogeneity $\eta(\vv{r})$ is obtained from (\ref{ot:reconst:reconstructedeta}).

\section{Simulation}
\label{siml}

To show that the numerical schemes developed in \S\ref{mrrf} and \S\ref{fn} are capable of optical tomography, we perform optical tomography with structured illumination. We consider a random medium which has the following optical properties.
\al{
\bar{\mu}_a=0.05\,{\rm cm}^{-1},\quad\mu_s=100\,{\rm cm}^{-1},\quad 
\mathrm{g}=0.9005.
\label{siml:opticalparameters}
}
For these optical parameters, the transport mean free path is $\ell^*=(\mu_t-\mu_sg)^{-1}=1\,{\rm mm}$, which is typical in biological tissue.

For the formulation in \S\ref{ot}, the forward data $\mathcal{D}(\vv{q}_0,\vv{q})$ is calculated by the diffusion equation, which is an approximation of the radiative transport equation. In this way, we can avoid inverse crime. We assume $N_a$ point targets at positions $\vv{r}_a^{(i)}=(\vv{\rho}_a^{(i)},z_a^{(i)})$ ($i=1,\dots,N_a$). Appendix \ref{fwd} is devoted to the computation of the data function $\mathcal{D}(\vv{q}_0,\vv{q})$.

We place detectors on a square lattice of spacing $h_d$. Detector positions are specified by
\eq{
(x_i,y_j)=(ih_d,jh_d),\quad i,j=-N_d,\dots,N_d.
}
We set
\eq{
h_d=1\,{\rm mm}\,,\qquad N_d=25.
}
Let $q^{(x)(i)},q^{(y)(j)}$ denote the $x$- and $y$-components of $\vv{q}$. We put $W=[2(N_d+N_s)+1]h_d$, and
\eq{
q^{(x)(i)}=\frac{2\pi}{W}i,\quad
q^{(y)(j)}=\frac{2\pi}{W}j,\quad
i,j=-N_d,\dots,N_d.
}
For $\vv{q}_0$ we use
\eq{
q_0^{(x)(i)}=\frac{2\pi}{W}i,\quad
q_0^{(y)(j)}=\frac{2\pi}{W}j,\quad
i,j=-N_s,\dots,N_s.
}
We set
\eq{
N_s=10.
}
Note that $q_0^{(x)(i+1)}-q_0^{(x)(i)}\approx 0.1\,{\rm mm}^{-1}$. In this demonstration we chose $L=N=9$, $I_0=1$, and $\uv_0=\hvv{z}$. Reconstruction is done by the formula (\ref{ot:reconst:reconstructedeta}). In both cases of MRRF and the three-dimensional $F_N$ method, about 10 singular values were used for the truncated SVD.

\subsection{A point absorber}

First we put $N_a=1$. Let us assume that a point absorber is embedded at $x_a^{(1)}=y_a^{(1)}=0$, $z_a^{(1)}=20\,{\rm mm}$.

\begin{figure}[ht]
\begin{center}
\includegraphics[width=0.8\textwidth]{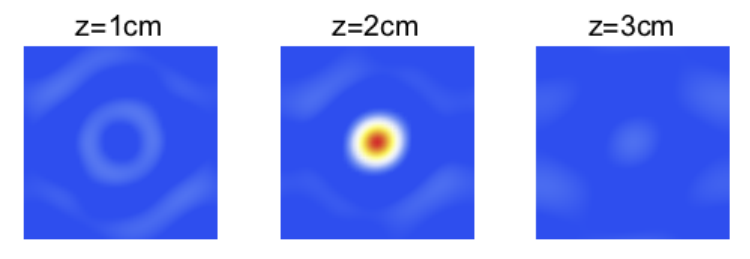}
\vspace{3mm} \\
\includegraphics[width=0.6\textwidth]{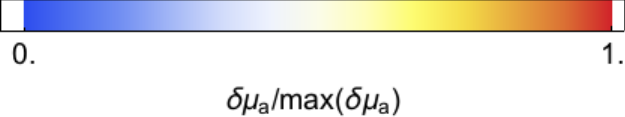}
\end{center}
\caption{
Reconstruction by MRRF. Three panels show the reconstruction of $\delta\mu_a(\vv{r})/\max_{\vv{r}\in\Rm^3_+}\left(\delta\mu_a(\vv{r})\right)$ in planes parallel to the $x$-$y$ plane at depths, from the left, $1\,{\rm cm}$, $2\,{\rm cm}$, and $3\,{\rm cm}$. The point absorber is placed at the center of the plane on the $z$-axis at the depth $2\,{\rm cm}$. The field of view is $5.1\,{\rm cm}\,\times\,5.1\,{\rm cm}$.
}
\label{fig1}
\end{figure}

Figure \ref{fig1} shows the reconstruction by MRRF. In Fig.~\ref{fig1}, $\delta\mu_a(\vv{r})/\max_{\vv{r}\in\Rm^3_+}\left(\delta\mu_a(\vv{r})\right)$ is plotted in planes parallel to the $x$-$y$ plane at different depths $z$. In the middle panel for $z=2\,{\rm cm}$, the absorber placed $2$ cm away from the surface is reconstructed while the reconstructed $\eta\approx0$ in the left panel for $z=1\,{\rm cm}$ and the right panel for $z=3\,{\rm cm}$.

\begin{figure}[ht]
\begin{center}
\includegraphics[width=0.8\textwidth]{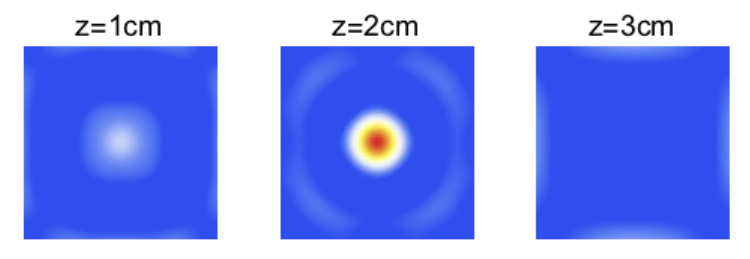}
\vspace{3mm} \\
\includegraphics[width=0.6\textwidth]{bar}
\end{center}
\caption{
Reconstruction by the three-dimensional $F_N$ method. Three panels show the reconstruction of $\delta\mu_a(\vv{r})/\max_{\vv{r}\in\Rm^3_+}\left(\delta\mu_a(\vv{r})\right)$ in planes parallel to the $x$-$y$ plane at depths, from the left, $1\,{\rm cm}$, $2\,{\rm cm}$, and $3\,{\rm cm}$. The point absorber is placed at the center of the plane on the $z$-axis at the depth $2\,{\rm cm}$. The field of view is $5.1\,{\rm cm}\,\times\,5.1\,{\rm cm}$.
}
\label{fig2}
\end{figure}

Figure \ref{fig2} shows the reconstruction by the three-dimensional $F_N$ method. The results are almost identical, and the target is successfully reconstructed.

\subsection{Two point absorbers}

\begin{figure}[ht]
\begin{center}
\includegraphics[width=0.8\textwidth]{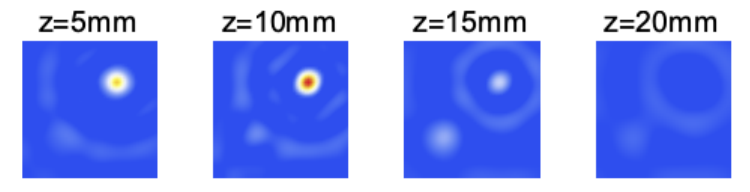}
\vspace{3mm} \\
\includegraphics[width=0.6\textwidth]{bar}
\end{center}
\caption{
Reconstruction by MRRF. Four panels show the reconstruction of $\delta\mu_a(\vv{r})/\max_{\vv{r}\in\Rm^3_+}\left(\delta\mu_a(\vv{r})\right)$ in planes parallel to the $x$-$y$ plane at depths, from the left, $5\,{\rm mm}$, $10\,{\rm mm}$, $15\,{\rm mm}$, and $20\,{\rm mm}$. The point absorbers are placed at $(10\,{\rm mm},\;10\,{\rm mm},\;10\,{\rm mm})$ and $(-10\,{\rm mm},\;-10\,{\rm mm},\;15\,{\rm mm})$. The field of view is $5.1\,{\rm cm}\,\times\,5.1\,{\rm cm}$.
}
\label{fig3}
\end{figure}

\begin{figure}[ht]
\begin{center}
\includegraphics[width=0.8\textwidth]{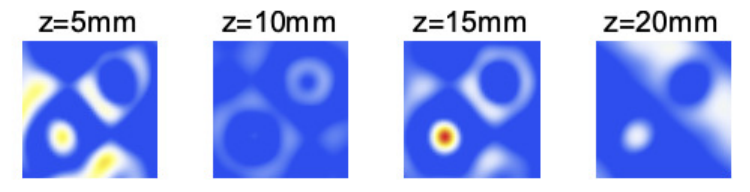}
\vspace{3mm} \\
\includegraphics[width=0.6\textwidth]{bar}
\end{center}
\caption{
Reconstruction by the three-dimensional $F_N$ method. Four panels show the reconstruction of $\delta\mu_a(\vv{r})/\max_{\vv{r}\in\Rm^3_+}\left(\delta\mu_a(\vv{r})\right)$ in planes parallel to the $x$-$y$ plane at depths, from the left, $5\,{\rm mm}$, $10\,{\rm mm}$, $15\,{\rm mm}$, and $20\,{\rm mm}$. The point absorbers are placed at $(10\,{\rm mm},\;10\,{\rm mm},\;10\,{\rm mm})$ and $(-10\,{\rm mm},\;-10\,{\rm mm},\;15\,{\rm mm})$. The field of view is $5.1\,{\rm cm}\,\times\,5.1\,{\rm cm}$.
}
\label{fig4}
\end{figure}

Next we consider two absorbers ($N_a=2$): One at $\vv{r}_a^{(1)}=(10\,{\rm mm},\;10\,{\rm mm},\;10\,{\rm mm})$ and the other at $\vv{r}_a^{(2)}=(-10\,{\rm mm},\;-10\,{\rm mm},\;15\,{\rm mm})$. Reconstructed images are shown in Figs.~\ref{fig3} and \ref{fig4} for MRRF and the three-dimensional $F_N$ method, respectively. Tomographic images show that reconstruction of multiple targets is more difficult than that of a single target. In Fig.~\ref{fig3} for MRRF, the target at $\vv{r}_a^{(2)}$ is almost invisible whereas the target at $\vv{r}_a^{(1)}$ is almost invisible in Fig.~\ref{fig4} for the three-dimensional $F_N$ method.

\begin{figure}[ht]
\begin{center}
\includegraphics[width=0.8\textwidth]{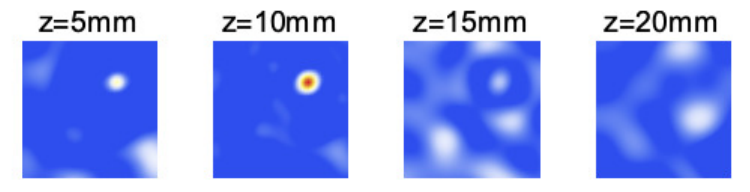}
\vspace{3mm} \\
\includegraphics[width=0.6\textwidth]{bar}
\end{center}
\caption{
Same as Fig.~\ref{fig3} but 3\% Gaussian noise is added.
}
\label{fig5}
\end{figure}

\begin{figure}[ht]
\begin{center}
\includegraphics[width=0.8\textwidth]{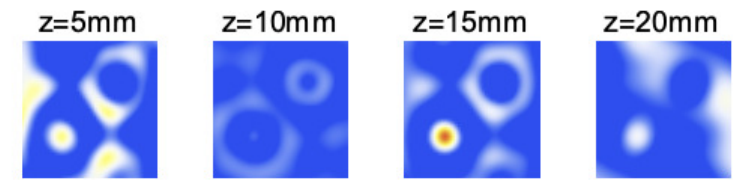}
\vspace{3mm} \\
\includegraphics[width=0.6\textwidth]{bar}
\end{center}
\caption{
Same as Fig.~\ref{fig4} but 3\% Gaussian noise is added.
}
\label{fig6}
\end{figure}

To investigate the robustness of reconstruction, we added 3\% Gaussian noise to the data function $\mathcal{D}(\vv{q}_0,\vv{q})$. The results are shown in Figs.~\ref{fig5} and \ref{fig6}.

\section{Concluding remarks}
\label{conclusions}

Since the aim of this paper is to present novel numerical algorithms of the radiative transport equation for optical tomography in spatial-frequency domain, we mainly focused on how to compute $\tilde{I}^{(1)},\tilde{I}^{(2)}$ in (\ref{DI1I2}) and reconstructions are done for simple point targets. Reconstructions of targets of more complicated shapes will be necessary in the future study. In particular, it is a future problem to show the superiority of the transport-based optical tomography over the diffusion-based optical tomography in spatial-frequency domain. Since structured illumination is quite often considered in shallow regions where the diffusion approximation breaks \cite{Konecky-etal09}, the radiative transport equation is expected to be important for structured illumination.

It is straightforward to extend the three-dimensional $F_N$ method for the half-space to the slab as is described in Appendix \ref{slab}. In \cite{Machida10}, MRRF was formulated in a slab. Thus our optical tomography can be similarly formulated in the slab geometry.

Although the first Born approximation is employed in this paper (see (\ref{born:firstBorn})), recently optical tomography with higher-order nonlinear terms has been developed by the use of a recursion algorithm for the inverse Born series of the radiative transport equation \cite{Machida-Schotland15}. The present schemes for calculating the kernel $K$ with MRRF and the three-dimensional $F_N$ method can readily be extended to such nonlinear inverse problems, which is an interesting future problem.

\section*{Acknowledgments}
The author acknowledges support from Grant-in-Aid for Scientific Research [17K05572 and 17H02081] of the Japan Society for the Promotion of Science (JSPS) and from the JSPS A3 foresight program: Modeling and Computation of Applied Inverse Problems. This work was also supported by Hamamatsu University School of Medicine [HUSM Grant-in-Aid].

\appendix

\section{The $F_N$ method in the slab geometry}
\label{slab}

We consider a slab of width $z_{\rm max}$. The radiative transport equation 
is given by 
\al{
\left\{\begin{aligned}
\uv\cdot\nabla I(\vv{r},\uv)+I(\vv{r},\uv)
=\varpi\int_{\Sm^2}p(\uv,\uv')I(\vv{r},\uv')\,d\uv',
\quad(\vv{r},\uv)\in\Rm^3_+\times\Sm^2,
\\
I(\vv{r},\uv)=f_1(\vv{\rho},\uv),
\quad z=0,\quad\vv{\rho}\in\Rm^2,\quad\uv\in\Sm^2_+,
\\
I(\vv{r},\uv)=f_2(\vv{\rho},\uv),
\quad z=z_{\rm max},\quad\vv{\rho}\in\Rm^2,\quad\uv\in\Sm^2_-,
\end{aligned}\right.
\label{slab:rte1}
}
with some boundary values $f_1(\vv{\rho},\uv),f_2(\vv{\rho},\uv)$. The specific intensity is given by
\eq{
&
I(\vv{r},\uv)
=\frac{1}{(2\pi)^2}\int_{\Rm^2}e^{i\vv{q}\cdot\vv{\rho}}
\sum_{m=-L}^L\Biggl[\sum_{j=0}^{M^m-1}A^m(\nu_j^m)
\rrf{\uvk(\nu_j^m,\vv{q})}\Phi_j^m(\uv)e^{-\hat{k}_z(\nu_j^mq)z/\nu_j^m}
\\
&+
\sum_{j=0}^{M^m-1}A^m(-\nu_j^m)\rrf{\uvk(-\nu_j^m,\vv{q})}
\Phi_{-j}^m(\uv)e^{\hat{k}_z(\nu_j^mq)z/\nu_j^m}
\\
&+
\int_{-1}^1A^m(\nu)\rrf{\uvk(\nu,\vv{q})}\Phi_{\nu}^m(\uv)
e^{-\hat{k}_z(\nu q)z/\nu}\,d\nu\Biggr],
}
with some coefficients $A^m(\pm\nu_j^m)$, $A^m(\nu)$. Using the orthogonality 
relations we have
\al{
\int_{\Sm^2}\mu\left(\rrf{\uvk}\Phi_{-\xi}^{m*}(\uv)\right)
I(\vv{\rho},0,\uv)\,d\uv=
2\pi\hat{k}_z(\xi q)\mathcal{N}(-\xi)A^m(-\xi),
\label{slab:proj1}
}
\al{
\int_{\Sm^2}\mu\left(\rrf{\uvk}\Phi_{-\xi}^{m*}(\uv)\right)
I(\vv{\rho},z_{\rm max},\uv)\,d\uv=
2\pi\hat{k}_z(\xi q)\mathcal{N}(-\xi)A^m(-\xi)
e^{\hat{k}_z(\xi q)z_{\rm max}/\xi},
\label{slab:proj2}
}
\al{
\int_{\Sm^2}\mu\left(\rrf{\uvk}\Phi_{\xi}^{m*}(\uv)\right)
I(\vv{\rho},0,\uv)\,d\uv=
2\pi\hat{k}_z(\xi q)\mathcal{N}(\xi)A^m(\xi),
\label{slab:proj3}
}
\al{
\int_{\Sm^2}\mu\left(\rrf{\uvk}\Phi_{\xi}^{m*}(\uv)\right)
I(\vv{\rho},z_{\rm max},\uv)\,d\uv=
2\pi\hat{k}_z(\xi q)\mathcal{N}(\xi)A^m(\xi)
e^{-\hat{k}_z(\xi q)z_{\rm max}/\xi},
\label{slab:proj4}
}
\al{
\int_{\Sm^2}\mu\left(\rrf{\uvk}\Phi_{-\xi}^{m*}(\uv)\right)
I(\vv{\rho},z,\uv)\,d\uv=
2\pi\hat{k}_z(\xi q)\mathcal{N}(-\xi)A^m(-\xi)e^{\hat{k}_z(\xi q)z/\xi},
\label{slab:proj5}
}
\al{
\int_{\Sm^2}\mu\left(\rrf{\uvk}\Phi_{\xi}^{m*}(\uv)\right)
I(\vv{\rho},z,\uv)\,d\uv=
2\pi\hat{k}_z(\xi q)\mathcal{N}(\xi)A^m(\xi)e^{-\hat{k}_z(\xi q)z/\xi}.
\label{slab:proj6}
}

We express $\tilde{I}(\vv{q},z,\uv)$ ($\uv\in\Sm^2$) as
\eq{
\tilde{I}(\vv{q},0,-\uv)&=
\sum_{m=-N}^N\sum_{\alpha=0\atop l=|m|+2\alpha}^{\lfloor(N-|m|)/2\rfloor}
c_{lm}(\vv{q})Y_{lm}(\uv),
\\
\tilde{I}(\vv{q},z_{\rm max},\uv)&=
\sum_{m=-N}^N\sum_{\alpha=0\atop l=|m|+2\alpha}^{\lfloor(N-|m|)/2\rfloor}
d_{lm}(\vv{q})Y_{lm}(\uv),
\\
\tilde{I}(\vv{q},z,-\uv)&=
\sum_{m=-N}^N\sum_{\alpha=0\atop l=|m|+2\alpha}^{\lfloor(N-|m|)/2\rfloor}
b_{lm}(\vv{q},z)Y_{lm}(\uv),
\\
\tilde{I}(\vv{q},z,\uv)&=
\sum_{m=-N}^N\sum_{\alpha=0\atop l=|m|+2\alpha}^{\lfloor(N-|m|)/2\rfloor}
a_{lm}(\vv{q},z)Y_{lm}(\uv),
}
where $\uv\in\Sm^2_+$. From (\ref{slab:proj1}) and (\ref{slab:proj2}) we obtain
\eq{
&
-(-1)^m\int_{\Sm^2_+}\mu\left(\rrf{\uvk}\Phi_{\xi}^{m*}(\uv)\right)
\tilde{I}(\vv{q},0,-\uv)\,d\uv+
\int_{\Sm^2_+}\mu\left(\rrf{\uvk}\Phi_{-\xi}^{m*}(\uv)\right)
\tilde{f}_1(\vv{q},\uv)\,d\uv
\\
&=
e^{-\hat{k}_z(\xi q)z_{\rm max}/\xi}\int_{\Sm^2_+}\mu
\left(\rrf{\uvk}\Phi_{-\xi}^{m*}(\uv)\right)\tilde{I}(\vv{q},z_{\rm max},\uv)
\,d\uv
\\
&-
(-1)^me^{-\hat{k}_z(\xi q)z_{\rm max}/\xi}\int_{\Sm^2_+}\mu
\left(\rrf{\uvk}\Phi_{\xi}^{m*}(\uv)\right)\tilde{f}_2(\vv{q},\uv)\,d\uv,
}
where we used $\Phi_{-\xi}^m(\uv)=(-1)^m\Phi_{\xi}^m(-\uv)$. Similarly from (\ref{slab:proj3}) and (\ref{slab:proj4}) we obtain
\eq{
&\int_{\Sm^2_+}\mu\left(\rrf{\uvk}\Phi_{\xi}^{m*}(\uv)\right)
\tilde{I}(\vv{q},z_{\rm max},\uv)\,d\uv-
(-1)^m\int_{\Sm^2_+}\mu\left(\rrf{\uvk}\Phi_{-\xi}^{m*}(\uv)\right)
\tilde{f}_2(\vv{q},\uv)\,d\uv
\\
&=
e^{-\hat{k}_z(\xi q)z_{\rm max}/\xi}\int_{\Sm^2_+}\mu
\left(\rrf{\uvk}\Phi_{\xi}^{m*}(\uv)\right)\tilde{f}_1(\vv{q},\uv)\,d\uv
\\
&-
(-1)^me^{-\hat{k}_z(\xi q)z_{\rm max}/\xi}\int_{\Sm^2_+}\mu
\left(\rrf{\uvk}\Phi_{-\xi}^{m*}(\uv)\right)\tilde{I}(\vv{q},0,-\uv)\,d\uv.
}
They are rewritten as
\al{
&
\sum_{lm'}\mathcal{B}_{lm'}^m(\xi)c_{lm'}+e^{-\hat{k}_z(\xi q)z_{\rm max}/\xi}
\sum_{lm'}\mathcal{A}_{lm'}^m(-\xi)d_{lm'}=
E_{12}^m,
\label{slab:key1}
\\
&
e^{-\hat{k}_z(\xi q)z_{\rm max}/\xi}\sum_{lm'}\mathcal{A}_{lm'}^m(-\xi)c_{lm'}
+\sum_{lm'}\mathcal{B}_{lm'}^m(\xi)d_{lm'}=
E_{21}^m,
\label{slab:key2}
}
where
\eq{
&
\mathcal{A}_{lm'}^m(-\xi)
=\int_{\Sm^2_+}\mu\left(\rrf{\uvk}\Phi_{-\xi}^{m*}(\uv)\right)
Y_{lm'}(\uv)\,d\uv,
\\
&
\mathcal{B}_{lm'}^m(\xi)
=(-1)^m\int_{\Sm^2_+}\mu\left(\rrf{\uvk}\Phi_{\xi}^{m*}(\uv)\right)
Y_{lm'}(\uv)\,d\uv,
\\
&
E_{12}^m
=\int_{\Sm^2_+}\mu\left(\rrf{\uvk}\Phi_{-\xi}^{m*}(\uv)\right)
\tilde{f}_1(\vv{q},\uv)\,d\uv
\\
&+
(-1)^me^{-\hat{k}_z(\xi q)z_{\rm max}/\xi}\int_{\Sm^2_+}\mu
\left(\rrf{\uvk}\Phi_{\xi}^{m*}(\uv)\right)\tilde{f}_2(\vv{q},\uv)\,d\uv,
\\
&
E_{21}^m
=\int_{\Sm^2_+}\mu\left(\rrf{\uvk}\Phi_{-\xi}^{m*}(\uv)\right)
\tilde{f}_2(\vv{q},\uv)\,d\uv
\\
&+
(-1)^me^{-\hat{k}_z(\xi q)z_{\rm max}/\xi}\int_{\Sm^2_+}\mu
\left(\rrf{\uvk}\Phi_{\xi}^{m*}(\uv)\right)\tilde{f}_1(\vv{q},\uv)\,d\uv.
}
We obtain by using (\ref{slab:proj1}) and (\ref{slab:proj5})
\eq{
\int_{\Sm^2}\mu\left(\rrf{\uvk}\Phi_{-\xi}^{m*}(\uv)\right)
\tilde{I}(\vv{q},0,\uv)\,d\uv=
e^{-\hat{k}_z(\xi q)z/\xi}\int_{\Sm^2}\mu
\left(\rrf{\uvk}\Phi_{-\xi}^{m*}(\uv)\right)\tilde{I}(\vv{q},z,\uv)\,d\uv,
}
from (\ref{slab:proj4}) and (\ref{slab:proj6})
\eq{
\int_{\Sm^2}\mu\left(\rrf{\uvk}\Phi_{\xi}^{m*}(\uv)\right)
\tilde{I}(\vv{q},z_{\rm max},\uv)\,d\uv=
e^{-\hat{k}_z(\xi q)(z_{\rm max}-z)/\xi}\int_{\Sm^2}\mu
\left(\rrf{\uvk}\Phi_{\xi}^{m*}(\uv)\right)\tilde{I}(\vv{q},z,\uv)\,d\uv,
}
by using (\ref{slab:proj2}) and (\ref{slab:proj5})
\al{
\int_{\Sm^2}\mu\left(\rrf{\uvk}\Phi_{-\xi}^{m*}(\uv)\right)
\tilde{I}(\vv{q},z,\uv)\,d\uv
=e^{-\hat{k}_z(\xi q)(z_{\rm max}-z)/\xi}
\int_{\Sm^2}\mu
\left(\rrf{\uvk}\Phi_{-\xi}^{m*}(\uv)\right)\tilde{I}(\vv{q},z_{\rm max},\uv)
\,d\uv,
\label{slab:from25}
}
and from (\ref{slab:proj3}) and (\ref{slab:proj6})
\al{
\int_{\Sm^2}\mu\left(\rrf{\uvk}\Phi_{\xi}^{m*}(\uv)\right)
\tilde{I}(\vv{q},z,\uv)\,d\uv=
e^{-\hat{k}_z(\xi q)z/\xi}\int_{\Sm^2}\mu
\left(\rrf{\uvk}\Phi_{\xi}^{m*}(\uv)\right)\tilde{I}(\vv{q},0,\uv)\,d\uv.
\label{slab:from36}
}

We obtain $c_{lm},d_{lm}$ from the linear system (\ref{slab:key1}), (\ref{slab:key2}). Then $a_{lm},b_{lm}$ are computed using (\ref{slab:from25}), (\ref{slab:from36}). We note that (\ref{slab:from25}) corresponds to (\ref{fn:proj5}) and (\ref{slab:from36}) corresponds to (\ref{fn:key}).

\section{Forward problem}
\label{fwd}

We assume point absorbers given by
\eq{
\eta(\vv{r})=\eta_aV_a\sum_{i=1}^{N_a}\delta(\vv{r}-\vv{r}_a^{(i)}),
}
where $V_a$ corresponds to the volume of each absorber and $\eta_a$ is a positive constant. We will calculate the data function using diffusion approximation. As is mentioned in (\ref{ellt}), the unit of length is $\ell_t=1/\mu_t$. With diffusion approximation we can express the Green's function as \cite{Markel-Schotland04}
\eq{
G(\vv{r},\uv;\vv{r}',\uv')
\simeq\frac{c\mu_s'\ell^*}{4\pi}(1+\ell^*\uv\cdot\nabla)
(1-\ell^*\uv'\cdot\nabla')G^{\rm (DE)}(\vv{r}',\vv{r}),
}
where $\mu_s'=\varpi(1-g)$ and $\ell^*=1/(1-\varpi g)$. Here,
\eq{
\left\{\begin{aligned}
-D_0\nabla^2G^{\rm (DE)}(\vv{r},\vv{r}')
+\alpha_0G^{\rm (DE)}(\vv{r},\vv{r}')=\delta(\vv{r}-\vv{r}'),
&\quad z>0,
\\
G-\ell\hvv{z}\cdot\nabla G=0,
&\quad z=0,
\end{aligned}\right.
}
where
\eq{
D_0=\frac{c\ell^*}{3},\qquad\alpha_0=c\bar{\mu}_a,
}
with $c$ the speed of light in the medium. We choose the extrapolation distance ($0\le\ell<\infty$) as 
\eq{
\ell=\frac{2}{3}\ell^*.
}
In the Fourier space we can introduce $\tilde{G}^{\rm (DE)}(z,z';\vv{q})=\tilde{G}^{\rm (DE)}(z,z';q)$ as
\al{
G^{\rm (DE)}(\vv{r},\vv{r}')
&=
\frac{1}{(2\pi)^2}\int_{\Rm^2}e^{i\vv{q}\cdot(\vv{\rho}-\vv{\rho}')}\tilde{G}^{\rm (DE)}(z,z';q)\,d\vv{q}
\nonumber \\
&=
\frac{1}{2\pi}\int_0^{\infty}qJ_0(q|\vv{\rho}-\vv{\rho}'|)\tilde{G}^{\rm (DE)}(z,z';q)\,dq,
\label{besint}
}
where $J_0$ is the Bessel function of the first kind of order $0$. We further approximate the Green's function as
\eq{
\tilde{G}(z,\uv;z',\uv';\vv{q})
\simeq
\frac{c\mu_s'\ell^*}{4\pi}(1-i\ell^*\uv\cdot\vv{q})(1-i\ell^*\uv'\cdot\vv{q})
\tilde{G}^{\rm (DE)}(z,z';\vv{q}).
}
We have \cite{Markel-Schotland02}
\eq{
\tilde{G}^{\rm (DE)}(z,z';q)
=\frac{1}{2D_0Q(q)}
\left[e^{-Q(q)|z-z'|}-\frac{1-Q(q)\ell}{1+Q(q)\ell}
e^{-Q(q)|z+z'|}\right],
}
where
\eq{
Q(q)&=\sqrt{k_0^2+q^2},
\\
k_0&=
\sqrt{\frac{\alpha_0}{D_0}}=\sqrt{\frac{3\bar{\mu}_a}{\ell^*}}
=\sqrt{3(1-\varpi)(1-\varpi g)}.
}
We have $k_0\approx0.12\,{\rm mm}^{-1}$ for optical parameters in (\ref{siml:opticalparameters}). In particular,
\eq{
\tilde{G}^{\rm (DE)}(z,0;q)
=\frac{\ell}{D_0(1+Q(q)\ell)}e^{-Q(q)z},\qquad z>0.
}
The integral in (\ref{besint}) can be numerically evaluated by the double-exponential formula \cite{Ogata-Sugihara98}. Let $0<j_1<j_2<\cdots$ be zeros of $J_0$. We define
\eq{
F(q)=\tilde{G}^{\rm (DE)}(z,z';q)q.
}
Then we have
\eq{
&
G^{\rm (DE)}(\vv{r},\vv{r}')=
\frac{1}{2\pi|\vv{\rho}-\vv{\rho}'|}\int_0^{\infty}J_0(x)F\left(\frac{x}{|\vv{\rho}-\vv{\rho}'|}\right)\,dx
\\
&\approx
\frac{1}{\pi|\vv{\rho}-\vv{\rho}'|}\sum_{k=1}^{N_k}J_0(q_k|\vv{\rho}-\vv{\rho}'|)F(q_k)\frac{\left.d\phi/dy\right|_{y=hj_k/\pi}}{j_k(J_1(j_k))^2},
}
where $h>0$ is a small number, $N_k$ is an integer, $\phi(y)=y\tanh(\frac{\pi}{2}\sinh{y})$, and
\eq{
q_k=\frac{\pi}{h|\vv{\rho}-\vv{\rho}'|}\phi\left(\frac{h}{\pi}j_k\right).
}

Using the radiative transport equation, the energy density $u^{(0)}(\vv{r})$ is obtained as
\eq{
u^{(0)}(\vv{r})
=\frac{1}{c}\int_{\Sm^2}I^{(0)}(\vv{r},\uv)\,d\uv.
}
Its Fourier transform has the form
\eq{
\tilde{u}^{(0)}(\vv{q},z)
&=
\frac{1}{c}\int_{\Sm^2}\int_{\Sm^2}\int_0^{\infty}
\tilde{G}(z,\uv;z',\uv';\vv{q})\mu'
\tilde{f}(\vv{q},\uv')\delta(z')\,dz'd\uv'd\uv,
}
where we used $\uv_0=\hvv{z}$ and (\ref{fourierf}). By the diffusion approximation we have
\eq{
\tilde{u}^{(0)}(\vv{q},z)
\simeq(2\pi)^2I_0\mu_s'\ell^*\delta(\vv{q}-\vv{q}_0)
\tilde{G}^{\rm (DE)}(z,0;\vv{q}_0).
}
The Fourier transform of the following $u^{(0)}$ is given by the right-hand side of the above equation.
\eq{
u^{(0)}(\vv{r})=\int_{\Rm^3_+}G^{\rm (DE)}(\vv{r},\vv{r}')S(\vv{r}')\,d\vv{r}',
}
with the source term
\eq{
S(\vv{r})=I_0\mu_s'\ell^*e^{i\vv{q}_0\cdot\vv{\rho}}\delta(z).
}
In the diffusion approximation, $u(\vv{r})=(1/c)\int_{\Sm^2}I(\vv{r},\uv)\,d\uv$, where $I(\vv{r},\uv)$ is the solution to (\ref{rte1}), satisfies
\eq{
\left\{\begin{aligned}
-D_0\nabla^2u(\vv{r})+(\alpha_0+c\eta(\vv{r}))u(\vv{r})=S(\vv{r}),
&\quad z>0,
\\
u-\ell\hvv{z}\cdot\nabla u=0,
&\quad z=0.
\end{aligned}\right.
}
We have
\al{
u(\vv{r})
&=
\int_{\Rm^3_+}G^{\rm (DE)}(\vv{r},\vv{r}')\left[S(\vv{r}')
-c\eta(\vv{r}')u(\vv{r}')\right]\,d\vv{r}'
\nonumber \\
&=
u^{(0)}(\vv{r})-c\int_{\Rm^3_+}G^{\rm (DE)}(\vv{r},\vv{r}')
\eta(\vv{r}')u(\vv{r}')\,d\vv{r}'
\nonumber \\
&=
u^{(0)}(\vv{r})-\gamma_0\sum_{i=1}^{N_a}G^{\rm (DE)}(\vv{r},\vv{r}_a^{(i)})u(\vv{r}_a^{(i)}),
\label{fwd:born0}
}
where
\eq{
\gamma_0=c\eta_aV_a.
}
Let us set
\eq{
\eta_a=0.0015,\quad V_a=10^{-6}\,{\rm mm}^3\,\mu_t^3.
}
We note that $((1/V_a)\int_{V_a}\mu_a(\vv{r})\,d\vv{r})/\bar{\mu}_a\approx4$ for each absorber. When $\vv{r}=\vv{r}_a^{(i)}$, we can write (\ref{fwd:born0}) as
\eq{
\sum_{j=1}^{N_a}M_{ij}u(\vv{r}_a^{(j)})=u^{(0)}(\vv{r}_a^{(i)}),
}
where
\eq{
M_{ij}=\delta_{ij}+\gamma_0G^{\rm (DE)}(\vv{r}_a^{(i)},\vv{r}_a^{(j)}).
}
Thus we have
\eq{
u(\vv{r})=
u^{(0)}(\vv{r})-\sum_{i,j=1}^{N_a}G^{\rm (DE)}(\vv{r},\vv{r}_a^{(i)})T_{ij}u^{(0)}(\vv{r}_a^{(j)}),
}
where $T_{ij}=\gamma_0M_{ij}^{-1}$. If $N_a=1$, we have
\eq{
u(\vv{r})=
u^{(0)}(\vv{r})-\gamma_rG^{\rm (DE)}(\vv{r},\vv{r}_a^{(1)})u^{(0)}(\vv{r}_a^{(1)}),
}
where we write \cite{FSS07}
\eq{
\gamma_r=T_{11}=\frac{\gamma_0}{1+\gamma_0G^{\rm (DE)}(\vv{r}_a^{(1)},\vv{r}_a^{(1)})}.
}
Let us consider $G^{\rm (DE)}(\vv{r}_a,\vv{r}_a)$, where $\vv{r}_a=\vv{r}_a^{(i)}$ ($i=1,\dots,N_a$). Introducing the ultraviolet cutoff $\Lambda_c$ ($k_0\Lambda_c\ll1$), we define the Green's function $G^{\rm (DE)}(\vv{r}_a,\vv{r}_a)$ as \cite{FSS07}
\eq{
&
G^{\rm (DE)}(\vv{r}_a,\vv{r}_a)
=\frac{1}{4\pi D_0}\left[\int_0^{2\pi/\Lambda_c}\frac{q}{Q(q)}\,dq+
\int_0^{\infty}\frac{q}{Q(q)}\frac{Q(q)\ell-1}{Q(q)\ell+1}e^{-2Q(q)z_a}\,dq
\right]
\\
&=
\frac{1}{4\pi D_0}\left[
Q\left(\frac{2\pi}{\Lambda_c}\right)-k_0+\frac{e^{-2k_0z_a}}{2z_a}-
\frac{2}{\ell}e^{2z_a/\ell}E_1\left(2z_a(k_0+\frac{1}{\ell})\right)\right],
}
where $E_1$ is the exponential integral defined by
\eq{
E_1(z)=\int_z^{\infty}\frac{e^{-t}}{t}\,dt.
}
In the numerical calculation we set
\eq{
\Lambda_c=V_a^{1/3}.
}

The hemispheric flux is obtained as
\eq{
J_+^{\rm (DE)}(\vv{\rho})
&=
\frac{-1}{4\pi}\int_0^{2\pi}\int_{-1}^0\mu\left[cu(\vv{\rho},0)
-3D_0\uv\cdot\nabla u(\vv{\rho},0)\right]\,d\mu d\va
\\
&=
\frac{c}{2}\left(\frac{1}{2}+\frac{D_0}{c\ell}\right)u(\vv{\rho},0)
=\frac{c}{2}u(\vv{\rho},0).
}
We then have
\eq{
J_+^{{\rm (DE)}(0)}(\vv{\rho})-J_+^{\rm (DE)}(\vv{\rho})=
\frac{c}{2}\left(u^{(0)}(\vv{\rho},0)-u(\vv{\rho},0)\right)=
\frac{c}{2}\sum_{i,j=1}^{N_a}G^{\rm (DE)}(\vv{\rho},0,\vv{r}_a^{(i)})T_{ij}u^{(0)}(\vv{r}_a^{(j)}).
}
The data function is thus calculated as
\eq{
\mathcal{D}(\vv{q}_0,\vv{q})
&=
\frac{h_d^2}{I_0}\sum_{\vv{\rho}}e^{-i(\vv{q}+\vv{q}_0)\cdot\vv{\rho}}
\left[J_+^{{\rm (DE)}(0)}(\vv{\rho})-J_+^{\rm (DE)}(\vv{\rho})\right]
\\
&=
\frac{c\mu_s'\ell^*}{2}\sum_{i,j=1}^{N_a}T_{ij}e^{-i\vv{q}\cdot\vv{\rho}_a^{(i)}}e^{-i\vv{q}_0\cdot(\vv{\rho}_a^{(i)}-\vv{\rho}_a^{(j)})}
\tilde{G}^{\rm (DE)}(0,z_a^{(i)};|\vv{q}+\vv{q}_0|)
\tilde{G}^{\rm (DE)}(z_a^{(j)},0;q_0).
}
When $N_a=1$, we have
\eq{
\mathcal{D}(\vv{q}_0,\vv{q})=
\frac{18\pi\eta_aV_a\mu_s'\ell^2}{4\pi\ell^*+3\eta_aV_ac_0}
e^{-i\vv{q}\cdot\vv{\rho}_a}
\frac{e^{-Q(q_0)z_a}}{1+Q(q_0)\ell}
\frac{e^{-Q(|\vv{q}+\vv{q}_0|)z_a}}{1+Q(|\vv{q}+\vv{q}_0|)\ell},
}
where
\eq{
c_0=
Q\left(\frac{2\pi}{\Lambda_c}\right)-k_0+\frac{e^{-2k_0z_a}}{2z_a}-
\frac{2}{\ell}e^{2z_a/\ell}E_1\left(2z_a(k_0+\frac{1}{\ell})\right).
}
We note that $c_0\simeq Q\left(\frac{2\pi}{\Lambda_c}\right)-k_0$ for large $z_a$.


\bibliography{myref}
\bibliographystyle{plain}

\end{document}